\documentclass[class=article, crop=false]{standalone}
\usepackage[subpreambles=true]{standalone}

\usepackage[subpreambles=true]{standalone}
\usepackage{import}
\usepackage{xr-hyper}

\usepackage[english]{babel}
\usepackage[a4paper,
  top=2cm,bottom=2cm,
  left=1.5cm,right=1.5cm,
  marginparwidth=1.75cm]{geometry}

\usepackage{authblk}

\usepackage[
  style=numeric-comp,
  sorting=none,
  backend=biber
]{biblatex}

\DeclareFieldFormat{pages}{#1}

\renewbibmacro*{volume+number+eid}{%
  \printfield[bold]{volume}%
  \setunit{\addcomma\space}%
  \printfield{pages}%
}

\renewbibmacro*{issue+date}{%
  \printtext[parens]{\printfield{year}}%
}

\DeclareBibliographyDriver{article}{%
  \usebibmacro{author}%
  \setunit{\labelnamepunct}\newblock
  \usebibmacro{title}%
  \newunit\newblock
  \printfield{journaltitle}%
  \setunit{\space}%
  \usebibmacro{volume+number+eid}%
  \setunit{\space}%
  \usebibmacro{issue+date}%
  \setunit{\addperiod\space}%
  \printfield{doi}%
  \finentry
}
\usepackage{csquotes}
\usepackage{amsmath}
\usepackage{amssymb}
\usepackage{mathtools}
\usepackage{bm}
\usepackage{braket}
\usepackage{siunitx}
\DeclareSIUnit\angstrom{\text{Å}}
\usepackage[version=4]{mhchem}

\usepackage{graphicx}
\usepackage{rotating}
\usepackage{quantikz}
\usepackage{adjustbox}
\usepackage{float}
\usepackage{multirow}
\usepackage{multicol}
\usepackage{calc}

\usepackage{titlesec}

\usepackage[colorlinks=true,allcolors=blue]{hyperref}
\usepackage{url}
\usepackage{orcidlink}

\newcommand\varpm{\mathbin{\vcenter{\hbox{%
  \oalign{\hfil$\scriptstyle+$\hfil\cr
          \noalign{\kern-.3ex}
          $\scriptscriptstyle({-})$\cr}%
}}}}
\newcommand\varmp{\mathbin{\vcenter{\hbox{%
  \oalign{\hfil$\scriptstyle-$\hfil\cr
          \noalign{\kern-.3ex}
          $\scriptscriptstyle({+})$\cr}%
}}}}

\newcommand{\supplementsetup}{
  \setcounter{figure}{0}
  \setcounter{table}{0}
  \setcounter{section}{0}
  \setcounter{equation}{0}

  \renewcommand{\figurename}{Supplementary Figure}
  \renewcommand{\tablename}{Supplementary Table}
  \renewcommand{\thefigure}{\arabic{figure}}
  \renewcommand{\thetable}{\arabic{table}}
  \renewcommand{\thesection}{\arabic{section}}
  \renewcommand{\theequation}{S\arabic{equation}} 

  \newcommand{\sectionname}{Supplementary Note}

  \titleformat{\section}
    {\normalfont\Large\bfseries}
    {\sectionname~\thesection:}
    {1em}
    {}
}

\title{\vspace{-2.0cm}Improved Strategies for Fermionic Quantum Simulation \\ with Global Interactions}

\author[1]{Thierry N. Kaldenbach\thanks{Corresponding author mail: \url{thierry.kaldenbach@dlr.de}}~}
\author[1]{Erik Schultheis}
\author[2]{Niklas Stewen}
\author[1]{Gabriel Breuil}
\affil[1]{Institute of Materials Research, German Aerospace Center (DLR), Cologne, Germany}
\affil[2]{Institute for Applied Physics, Technical University of Darmstadt, Darmstadt, Germany}

\date{March 27, 2026}

\begin{document}
\maketitle

\vspace{-0.7cm}

\begin{abstract}
\begin{center} 
\begin{minipage}{0.8\textwidth}  
We present efficient quantum circuits for fermionic excitation operators tailored for ion trap quantum computers exhibiting the M\o{}lmer-S\o{}rensen (MS) gate. Such operators commonly arise in the study of static and dynamic properties in electronic structure problems using Unitary Coupled Cluster theory or Trotterized time evolution.
We detail how the global MS interaction naturally suits the non-local structure of fermionic excitation operators under the Jordan-Wigner mapping and simultaneously provides optimal parallelism in their circuit decompositions. Compared to previous schemes on ion traps, 
our approach reduces the number of MS gates by factors of 2-, and 4, for single-, and double excitations, respectively. These improvements promise significant speedups and error reductions, which we demonstrate by characterizing our circuits under a realistic pulse-level noise model of a linear ion trap quantum processor. \\ \\
\textit{npj Quantum Inf} DOI: \href{https://doi.org/10.1038/s41534-026-01223-0}{10.1038/s41534-026-01223-0} \hfill
\textit{arXiv} DOI: \href{https://doi.org/10.48550/arXiv.2504.03237}{10.48550/arXiv.2504.03237}
\end{minipage} 
\end{center}
\end{abstract}

\vspace{0.1cm}

\begin{multicols*}{2}

\section*{Introduction} \label{sec:Introduction}

Among various expected use-cases of quantum computation, digital quantum simulation of fermionic many-body systems stands out as one of the most promising prospects \cite{kandala_hardware_2017,malley_scalable_2016, McArdle2020Quantum}. Quantum simulations of electronic structure problems \cite{liu_prospects_2022} are expected to yield unprecedented insight in fields ranging from quantum chemistry 
to materials science and engineering 
or drug discovery \cite{magann_digital_2021,ollitrault_molecular_2021,clinton2024towards,wang_recent_2023}
. This expectation stems from the capability of quantum computers to exhibit superposition and entanglement, thus efficiently storing a combinatorially large number of electronic configurations, which is the bottleneck of many classical methods \cite{kandala_hardware_2017,malley_scalable_2016,liu_prospects_2022}. 

Electronic structure problems are typically mapped to quantum computers using a fermion-to-qubit mapping. In this formalism, the state of the system is encoded as a multi-qubit state and the Hamiltonian governing the problem is encoded as a weighted sum of Pauli operators. A large focus on fermionic mappings is dedicated to the optimization of mappings towards limited connectivity devices, where typically only interactions of one or two qubits are possible. One of the most popular approaches, the Jordan-Wigner (JW) transformation \cite{jordan1993paulische}, is highly limited in its applicability on such devices due to its linear 
Pauli weight scaling. More sophisticated mappings can be used to tackle this obstacle, e.g., the Bravyi-Kitaev (BK) mapping \cite{bravyi2002fermionic, seeley2012bravyi} which achieves
logarithmic localities. However, in practice, the benefit of logarithmic Pauli weight scaling is mitigated due to the need for many SWAP gates in the transpilation for a limited hardware connectivity \cite{Miller2023Bonsai}. Among numerous other approaches \cite{Chiew_2023, setia2018bravyi, Steudtner_2018, PhysRevA.95.032332, PhysRevA.99.022308}, tree-based mappings have recently proven to be particularly effective at simultaneously mitigating the Pauli weight and number of SWAP gates for specific connectivities \cite{Miller2023Bonsai, miller2024treespilation}.

The necessity for SWAP gates vanishes if one instead assumes a quantum device offering up-to-global interactions.
Such interactions are provided on ion trap simulators \cite{benhelm2008towards, taylor2017study} featuring the M\o{}lmer-S\o{}rensen (MS) gate \cite{sorensen1999quantum, Sorensen2000Entanglement}, which can be used to efficiently implement non-local Pauli rotations arising under the chosen fermionic mapping. Most importantly, any Pauli rotation can be implemented using two MS gates regardless of the underlying locality \cite{Müller2011Simulating}. In the context of fermionic systems, simulations leveraging the MS gate using the JW or BK mapping have been studied for dynamics in lattice models \cite{casanova2012quantum, lamata2014efficient, martinez2016real} and ground state computations in quantum chemistry \cite{Hempel2018Quantum, Romero2019Strategies, Shen2017Quantum} based on Unitary Coupled Cluster (UCC) theory \cite{bartlett1989alternative, peruzzo2014variational, anand2022quantum}. 

The task of implementing arbitrary quantum circuits in terms of MS gates has been studied in Refs.~\cite{Maslov_2018, Wetering_2021}. While Ref.~\cite{Wetering_2021} already provides tight bounds on the number of MS gates for generic circuits, their algorithm gets outperformed by handcrafted results for specific unitaries \cite{Ivanov2015Efficient, Maslov_2018, Groenland_2020}. The schemes presented in our work are specific to classes of unitaries in fermionic systems.  

\begin{figure*}[t]
    \centering
    \includegraphics[width=0.6\textwidth]{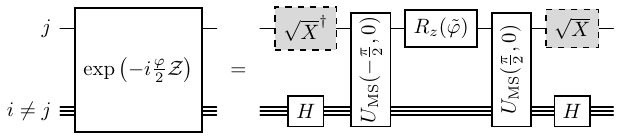}
    \caption{Circuit decomposition of the global rotation $U(\varphi)=\exp(-i\varphi/2\mathcal Z)$ using the \textbf{XX} gate. The rotation angle in the circuit is defined as $\tilde \varphi = (-1)^m\varphi$, where $m$ follows the distinction between even qubit numbers $n=2m$ and odd numbers $n=2m+1$ from equation~\ref{eq:Generator}. Gates with dashed lines are only required if $n$ is even to turn $Y_j$ into $Z_j$.}
    \label{fig:PauliRotation}
\end{figure*}

In this work, we show how the MS gate naturally implements the Pauli operator pool of fermionic and qubit excitation operators with maximum parallelism.
Our approach exploits that specific types of MS gates perform simultaneous diagonalization of certain Pauli operators arising for excitation operators under the JW transformation. 
Using this feature, we leverage previous works, where each non-local Pauli operator is realized by its own pair of MS gates \cite{Müller2011Simulating, casanova2012quantum, lamata2014efficient, Romero2019Strategies}, and achieve an MS gate reduction by a factor of 2 for quadratic terms, and a factor of 4 for quartic terms. 
Our technique is also ancilla-free, making it not only faster, but also cheaper in terms of qubit requirements. 
By exploiting the local fermionic equivalences between (anti-)symmetrized excitation operators, we can use our circuits as building blocks for both UCC calculations, as well as the time evolution of electronic structure Hamiltonians in second quantization \cite{berry2020time}. 
This enables the study of mixed quantum-classical dynamics within the Born-Oppenheimer approximation,
thus providing an hybrid framework for studying time-dependent properties in molecules \cite{lee_variational_2022,bultrini_mixed_2023,ollitrault_molecular_2021,berry2020time}.
After introducing the fermionic building blocks, we explicitly outline our techniques at hand of the \ce{H3+} molecule by showing how UCCSD-, and time evolution circuits can be efficiently assembled. Finally, we demonstrate the efficiency of our circuit decompositions by characterizing the circuits via noisy simulations of molecular ground states of various molecules on an $12$-qubit ion trap emulator at the pulse level.\\

Before presenting the results, we provide short introductions into general digital quantum simulations with MS gates, and the classes of fermionic operators used in UCCSD and Hamiltonian simulation. Readers with strong familiarity with those subjects are encouraged to skip to the results. 
\\

We first introduce the core properties of the MS gate and how to employ it to implement arbitrary Pauli rotations. For now, it is instructive to treat the MS gate as an idealized theoretical building block for quantum circuits, while an experimental description of the MS gate and its experimental challenges is later introduced in the 
\hyperref[sec:noise_modelling]{results} section.
The MS gate captures all pairwise two-qubit interactions and is parametrized by the two parameters $\theta$ and $\phi$,
\begin{equation}
    U_\text{MS}(\theta,\phi) = \exp\left[-i\frac{\theta}{4}\Bigl(\cos(\phi)S_x + \sin(\phi)S_y\Bigr)^2\right].
    \label{eq:MSGate}
\end{equation}
Here, $\theta$ is the phase and $\phi$ determines the type of interaction. The collective spin operators $S_x$ and $S_y$ are defined as the sum over all $n$ qubits involved in the gate, e.g.,~$S_x=\sum_{i=1}^n X_i$. Through the course of this work, we are concerned with two special cases where the MS gate is a non-identity Clifford operation, namely the $XX$- and $YY$-type interactions:
\begin{alignat}{2}
    \textbf{XX} &\coloneqq U_\text{MS}\left(\frac{\pi}{2}, 0\right) &&= \exp\Bigl(-i\frac{\pi}{4}\sum_{\substack{j < k}}X_j X_k\Bigr),
     \label{eq:MSGateXX} \\
    \textbf{YY} &\coloneqq U_\text{MS}\left(\frac{\pi}{2}, \frac{\pi}{2}\right) &&= \exp\Bigl(-i\frac{\pi}{4}\sum_{\substack{j < k}}Y_j Y_k\Bigr).
     \label{eq:MSGateYY}
\end{alignat}
The inverse gates $\textbf{XX}^\dagger$ and $\textbf{YY}^\dagger$ are obtained with $\theta=-\pi/2$, and due to the negative sign of $\theta$ sometimes referred to as \enquote{Backward} MS gates \cite{Müller2011Simulating}. Experimentally speaking, sign changes of $\theta$ are inconvenient since they require frequency changes of the driving field \cite{Müller2011Simulating}. This issue can be addressed by exploiting the local unitary equivalence between forward and backward MS gates, as detailed in the \hyperref[app:BackwardMS]{methods} section. 
However, for the sake of a compact circuit notation, we use both the forward- and backward MS gates for our circuits in this work. 

While the MS gate in equation~\ref{eq:MSGate} is globally defined, we typically do not want all qubits to interact at once. Instead, we need targeted MS gates acting on problem-specific subsets of qubits. 
From an experimental point of view, numerous approaches exist to restrict the MS interaction to subsets of the qubit array \cite{Müller2011Simulating, Martinez_2016, debnath2016demonstration, figgatt2019parallel, grzesiak2020efficient}. 
Alternatively, ions can be effectively decoupled by interspersing global MS gates with single-qubit gates \cite{Nebendahl2009Optimal, Wetering_2021}. Through this work, we make use of targeted MS gates, with the assumed experimental realization later outlined in the \hyperref[sec:noise_modelling]{results} section. 


We now outline how any unitary Pauli rotation $U=\exp(-i\varphi/2\mathcal{P})$, where $\mathcal{P}$ is an $N$-qubit Pauli string $\mathcal{P}\in\{I, X, Y, Z\}^{\otimes N}$, is decomposed into a sequence of three gate operations up to local Clifford transformations -- namely two MS gates and one local parameterized rotation. We mostly follow the same derivation as in Ref.~\cite{Müller2011Simulating}, however with an ancilla-free approach. For that purpose, we assume that $\mathcal{P}$ is $n$-local (with $n\leq N$)
and, w.l.o.g.~always acts non-trivially on some qubit $j$. Let us consider the unitary operator
\begin{equation}
    U^{(j)}(\varphi) = \textbf{XX} R_z^{(j)}(\varphi) \textbf{XX}^\dagger,
    \label{eq:PauliDecomposition_STEP_0}
\end{equation}
where \textbf{XX} acts on all $n$ qubits affected by the Pauli string $\mathcal{P}$, and $R_z^{(j)}(\varphi)=\exp(-i\varphi/2 Z_j)$ is the single-qubit $Z$-rotation gate acting on qubit $j$.
Since \textbf{XX} is Clifford, we may rewrite equation~\ref{eq:PauliDecomposition_STEP_0} as
\begin{equation}
    U^{(j)}(\varphi)=\exp\left(-i\frac{\varphi}{2}\mathcal{P}^{(j)}\right),
\end{equation}
where the generating Pauli string is given by $\mathcal{P}^{(j)}=\textbf{XX} Z_j\textbf{XX}^\dagger$. 
The structure of $\mathcal{P}^{(j)}$ is intrinsically linked to the locality $n$ of the MS gate and the qubit $j$ on which the $R_z$ rotation is carried out, 
\begin{equation}
    \mathcal{P}^{(j)} = \mathcal{X}^{(j)} \otimes
    \begin{cases}
        (-1)^m Y_j, &\text{for } n=2m, \\
        (-1)^m Z_j, &\text{for } n=2m+1, \\
    \end{cases}
    \label{eq:Generator}
\end{equation}
where $\mathcal{X}^{(j)} =\otimes_{i\neq j}X_i$ and $m\in \mathbb{N}$. For the proof, refer to Supplementary Note \hyperref[app:Generator]{1}.
Any $n$-local Pauli string $\mathcal{P}$ either directly assumes the form in equation~\ref{eq:Generator} through a suitable choice of $j$ ($n$ choices), or can be adjusted through local Clifford transformations accordingly. By instead  using $YY$-type interactions in equation~\ref{eq:PauliDecomposition_STEP_0}, one reverses the roles of $X$ and $Y$ in equation~\ref{eq:Generator}. 
This proves to be particularly convenient for the string pool in double excitations (cf.~\hyperref[subsubsec:DoubleExcitation]{results}).

In Fig.~\ref{fig:PauliRotation}, we show how the \textbf{XX} gate is used in a quantum circuit to achieve an $n$-qubit Pauli-$Z$ rotation, i.e.~a rotation generated by $\mathcal{Z}=\otimes_{i=1}^n Z_i$. We compensate for the different cases in equation~\ref{eq:Generator} by adjusting the rotation angle and/or adding local Cliffords based on the identity $\sqrt{X} Y \sqrt{X}^\dagger = Z$.

When instead decomposing an $n$-local Pauli rotation in terms of CNOT gates, a total number of $2(n-1)$ CNOTs is needed. Assuming full connectivity, these CNOTs could be arranged in $\mathcal{O}(\log(n))$ depth. In practice, however, this property can hardly be utilized due to the SWAP overhead.  Meanwhile, the number of MS gates remains constant at 2. It should be emphasized that the MS gate time also grows with the with the number of interacting qubits. Hence, despite a constant gate count, the increase in locality is still not for free. \\


Having set the scene for arbitrary quantum simulations with MS gates, we now explore the class of fermionic operations subject to this work.
The Unitary Coupled Cluster (UCC) ansatz is of particular interest due to its preservation of symmetries in electronic systems \cite{gard2020efficient}, such as the total particle number or the spin. In its most general form, it is defined as 
\begin{equation}
    \ket{\psi} = \exp\Bigl(\sum_N T_N\Bigr) \ket{\psi_0},
\end{equation}
where $\ket{\psi_0}$ is an initial guess of the systems ground state -- typically the Hartree-Fock ground state -- and $T_N$ denotes the $N$-th cluster operator incorporating all possible excitations of $N$ electrons from occupied to virtual orbitals. In practice, $N$ is often truncated at $2$, giving rise to the UCCSD ansatz $\exp(i(T_1+T_2))$, where the first- and second cluster operators
\begin{align}
        T_1&=\sum_{\substack{q\in\text{virt.}\\p\in\text{occ.}}}\theta_p^q G_p^q, \quad \text{and} \quad T_2&=\sum_{\substack{r,s\in\text{virt.}\\p,q\in\text{occ.}}}\theta_{pq}^{rs} G_{pq}^{rs}
        \label{eq:ClusterOperators}
\end{align}
entail all possible generators of single- and double excitation generators, which are defined by the antisymmetrized terms 
\begin{align}
    G_{p}^{q} &= i(a^\dagger_p a_q - \text{H.c.}), \label{eq:SingleExcitationGenerator}\\
    G_{pq}^{rs} &= i(a^\dagger_p a^\dagger_q a_r a_s - \text{H.c.}), \label{eq:DoubleExcitationGenerator}
\end{align}
respectively.
Next, $\exp(T_1+T_2)$ is typically approximated through a first-order Trotter-Suzuki product decomposition \cite{trotter1959product, suzuki1976generalized}, such that the ansatz is a sequence of the single- and double excitation operators
\begin{equation}
    U(\bm \theta) = \prod_{\substack{q\in\text{virt.}\\p\in\text{occ.}}} U_p^q(\theta_p^q) \prod_{\substack{r, s\in\text{virt.}\\p, q\in\text{occ.}}} U_{pq}^{rs}(\theta_{pq}^{rs}),
    \label{eq:UCCSDAnsatz}
\end{equation}
where 
\begin{align}
    U_p^q(\theta) = \exp(-i\theta G_p^q) \label{eq:SingleExcitation}, \\
    U_{pq}^{rs}(\theta)=\exp(-i\theta G_{pq}^{rs}) \label{eq:DoubleExcitation}
\end{align}
are the single- and double excitation operators, respectively.
The challenge of UCC(SD) then lies in determining the parameters $\bm \theta$ to minimize the variational energy $E(\bm \theta)=\braket{\psi_0|U^\dagger(\bm \theta) H U(\bm \theta)|\psi_0}$. Cost-efficient updating schemes exploiting the spectral properties of excitations are detailed in Refs.~\cite{kottmann2021Feasible, jaeger2024fast}. Our work deals with the efficient circuit decomposition of excitations and is compatible with these parameter optimization schemes.
\\

\begin{figure*}[t]
    \centering
    \includegraphics[width=\textwidth]{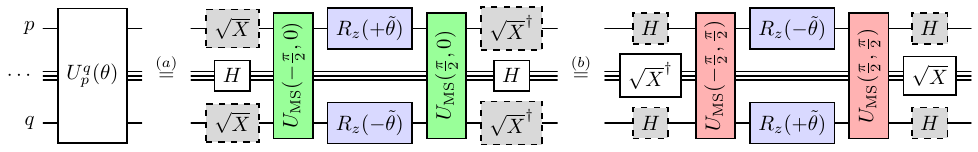}
    \caption{Circuit decomposition of the single-excitation gate $U_p^q(\theta)=\exp(-i\theta G_p^q)$ using (a) the \textbf{XX} gate and (b) the \textbf{YY} gate. The rotation angle in the circuits is defined as $\tilde \theta = (-1)^m\theta$, where $m$ follows the same distinction between even qubit numbers $n=2m$ and odd numbers $n=2m+1$ as before. Gates with dashed lines are only required if $n$ is odd. The dots $\cdots$ labeling the quantum wire bundle represent all qubits affected by the parity string $\mathcal{Z}_p^q$.}
    \label{fig:SingleExcitation}
\end{figure*}

To study the non-adiabatic dynamics of quantum many-body systems, the time-dependent Schrödinger equation has to be solved. The solution is given by the time evolution operator, which is generated by the electronic structure Hamiltonian describing the fermionic many-body system. 
The time-dependent electronic structure Hamiltonian
can be expressed in terms of the second-quantized operators and is defined as 
\begin{equation}
    H_{\mathrm{el.}}=\sum_{pq}h_{pq}a_p^{\dagger}a_{q} + \frac{1}{2} \sum_{pqrs}h_{pqrs}a_{p}^{\dagger}a_{q}^{\dagger}a_{r}a_{s},
    \label{eq:ham}
\end{equation}
where the one- and two-electron integrals $h_{pq}$ and $h_{pqrs}$ are subject to the permutational symmetries $h_{pq} = h^*_{qp}$ and $h_{pqrs} = h_{qpsr} = h^*_{rspq} = h^*_{srqp}$ \cite{szabo_modern_1996}. More information on these integrals is provided in Supplementary Note \hyperref[app:ElectronicHamiltonian]{2}. Here, we drop the explicit time dependence to shorten the notation. However, one could consider an explicit time dependence in the electron integrals due to dynamics in the nuclear coordinates. If one further assumes a time-dependent basis set, the fermionic operators would be time-dependent as well.
We exploit these symmetries to rewrite the electronic Hamiltonian as $H_{\mathrm{el.}} = H + \tilde H$, with the antisymmetrized Hamiltonian $H$ and symmetrized Hamiltonian $\tilde H$
\begin{align}
    H &= \frac{1}{2}\sum_{pq} \Im(h_{pq}) G_{p}^{q} + \frac{1}{4} \sum_{pqrs}\Im(h_{pqrs}) G_{pq}^{rs}, \label{eq:AntisymmetrizedHamiltonian} \\
    \tilde H &= \frac{1}{2} \sum_{pq} \Re(h_{pq}) \tilde G_{p}^{q} + \frac{1}{4} \sum_{pqrs} \Re(h_{pqrs}) \tilde G_{pq}^{rs}, \label{eq:SymmetrizedHamiltonian} 
\end{align}
where $\Re (\cdot)$ and $\Im (\cdot)$ denote the real- and imaginary parts, respectively. The symmetrized Hermitian generators $\tilde G$ are given by
\begin{align}
    \tilde G_{p}^{q} &= a^\dagger_p a_q + \text{H.c.}, \label{eq:SingleExcitationSymm}\\
    \tilde G_{pq}^{rs} &= a^\dagger_p a^\dagger_q a_r a_s + \text{H.c.}, \label{eq:DoubleExcitationSymm}
\end{align}
while the antisymmetrized Hermitian generators $G$ are the same as for the excitations in equations~\ref{eq:SingleExcitationGenerator} and \ref{eq:DoubleExcitationGenerator}. For a detailed derivation, we refer the reader to Supplementary Note \hyperref[app:ComplexOrbitals]{3}. 

The time evolution of the electronic structure system is governed by the unitary time evolution operator $U(t, t_0)=\exp_\mathcal{T}(-i\int_{t_0}^t d\tau H_\text{el.}(\tau))$, where $\exp_\mathcal{T}$ denotes the time ordered operator exponential. For a small time step $\delta t=t-t_0$, we may approximate $U(t,t_0)$ through a first-order Trotter-Suzuki product formula
\begin{align}
    U(\delta t) &= \prod_{pq} U_p^q(\Im(h_{pq})\delta t)\prod_{pqrs} U_{pq}^{rs}(\Im(h_{pqrs})\delta t) \nonumber \\
    &\times \prod_{pq}\tilde U_p^q(\Re(h_{pq})\delta t) \prod_{pqrs} \tilde U_{pq}^{rs}(\Re(h_{pqrs})\delta t),
\end{align}
where $U_p^q$ and $U_{pq}^{rs}$ are the single- and double excitations from equations~\ref{eq:SingleExcitation} and \ref{eq:DoubleExcitation}, and $\tilde U_p^q$ and $\tilde U_{pq}^{rs}$ are analogously defined with the symmetrized generators from equations~\ref{eq:SingleExcitationSymm} and \ref{eq:DoubleExcitationSymm}. We want to highlight that the unitaries corresponding to the antisymmetrized Hamiltonian $\prod_{pq} U_p^q\prod_{pqrs} U_{pq}^{rs}$ are structurally similar to the UCCSD ansatz in equation~\ref{eq:UCCSDAnsatz}, with the only differences being that no distinction between occupied and virtual orbitals is made and that shared indices (e.g., controlled excitations) are included.

\section*{Results}
In this section, we first present our optimized circuits for antisymmetrized fermionic single- and double excitations. We then outline how our procedure generalizes to arbitrary excitation orders. We further discuss the applicability to other fermion-to-qubit mappings, and show how our techniques readily apply to qubit-excitations. We then show how the symmetrized fermionic terms for Hamiltonian simulation can be recovered from the previous circuits. We then give explicit circuit examples for UCCSD ansätze and Trotter steps. Last, we introduce a realistic pulse-level noise model, which we use to assess the practical improvements of our circuits in noisy scenarios.  

\subsection*{Circuit for Single Excitations} \label{subsec:SingleExcitation}

Under the JW mapping (cf.~\hyperref[sec:JW_map]{methods}),
the generator of a single excitation between two orbitals $p, q$ with $p<q$ assumes the Pauli decomposition
\begin{equation}
    G_p^q \to \frac{1}{2}\mathcal{Z}_p^q \left(Y_p  X_q - X_p Y_q\right), \label{eq:SingleExcitationGeneratorPauli}
\end{equation}
with the parity string $\mathcal{Z}_p^q\coloneqq\prod_{j\in\{p, q\}}\bigotimes_{k<j}Z_k$. We now reproduce equation~\ref{eq:SingleExcitationGeneratorPauli} in terms of local operators and MS gates based on equation~\ref{eq:Generator} to infer the circuit decomposition of $U_p^q(\theta)$.
Since each Pauli string in the single excitation generator from equation~\ref{eq:SingleExcitationGeneratorPauli} consists of one $X$ and $Y$, the choice of either the \textbf{XX} or \textbf{YY} gate is arbitrary. For the sake of simplicity, we only consider \textbf{XX} here. 

We assume that the MS gate acts on all $n=q-p+1$ qubits affected by the single excitation. The core idea is that we can realize up to $n$ Pauli rotations in parallel by interspersing two MS gates with $R_z$ gates. The generator of the single excitation entails two Pauli strings always acting at least on the two qubits $p$ and $q$, therefore we use these two qubits to place the $R_z$ gates in parallel.

For an even number of qubits $n=2m$, we use
\begin{equation}
    \textbf{XX} (Z_p - Z_q) \textbf{XX}^\dagger = (-1)^m \mathcal{X}_p^q (Y_p X_q - X_p Y_q),
    \label{eq:SingleExcitationGeneratorDecompositionEven}
\end{equation}
where $\mathcal{X}$ is analogously defined to $\mathcal{Z}$ with Pauli-$X$ instead.
Note that this is already local Clifford equivalent to equation~\ref{eq:SingleExcitationGeneratorPauli} up to a Hadamard transformation on the qubits $p+1,\dots,q-1$ and a prefactor of $(-1)^m/2$. 
For an odd number of qubits $n=2m+1$, we find
\begin{align}
    \textbf{XX} (Z_p - Z_q) \textbf{XX}^\dagger = (-1)^m \mathcal{X}_{p}^{q} (Z_p X_q - X_p Z_q).
\end{align}
Here, we obtain $Z$ instead of $Y$. We circumvent that by exploiting that $\sqrt{X}^\dagger Z\sqrt{X}=Y$. By using this transformation on qubits $p$ and $q$, we can change $Z\to Y$ without affecting $X$. Overall, this gives rise to
\begin{align}
    &\sqrt{X}^\dagger_p\sqrt{X}^\dagger_q\textbf{XX} (Z_p - Z_q) \textbf{XX}^\dagger \sqrt{X}_p\sqrt{X}_q \label{eq:SingleExcitationGeneratorDecompositionOdd}\\
    &= \textbf{XX} (Y_p - Y_q) \textbf{XX}^\dagger = (-1)^m \mathcal{X}_p^q (Y_p X_q - X_p Y_q)
    \nonumber
\end{align}
The prefactor of $(-1)^m$ can in theory be absorbed into the variational parameter $\theta$ and thus be ignored in the circuit decomposition. However, we keep track of it as it becomes crucial for time evolution. 
Equations \ref{eq:SingleExcitationGeneratorDecompositionEven} and \ref{eq:SingleExcitationGeneratorDecompositionOdd} give rise to the circuit decompositions of a single excitation with \textbf{XX} gates depicted in Fig.~\ref{fig:SingleExcitation}(a). An equivalent decomposition in terms of \textbf{YY} gates is provided in Fig.~\ref{fig:SingleExcitation}(b). In Ref.~\cite{casanova2012quantum}, where every Pauli string is implemented with its own pair of MS gates, a total of 4 MS gates is required. With our parallelization, we achieve the same operation with only 2 MS gates, which is optimal with respect to the assumed gate set of targeted Clifford MS gates and arbitrary local unitaries.

\subsection*{Circuit for Double Excitations} \label{subsubsec:DoubleExcitation}

\begin{figure*}[t]
    \centering
    \includegraphics[width=0.85\textwidth]{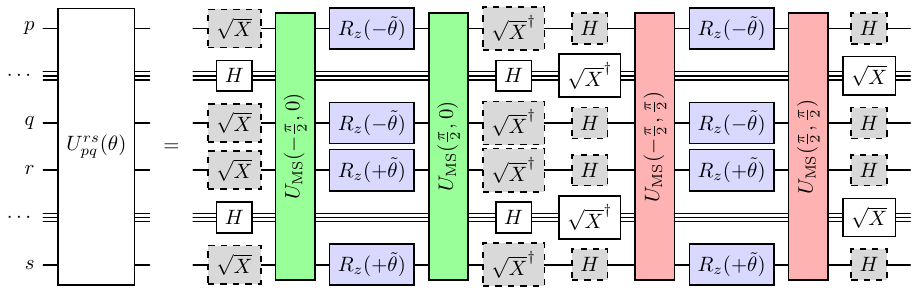}
    \caption{Circuit decomposition of the double-excitation gate $U_{pq}^{rs}(\theta)=\exp(-i\theta G_{pq}^{rs})$ using the \textbf{XX} and \textbf{YY} gates. The rotation angle in the circuits is defined as $\tilde \theta = (-1)^m\theta/4$, where $m$ follows the same distinction between even qubit numbers $n=2m$ and odd numbers $n=2m+1$ as before. Gates with dashed lines are only required if $n$ is odd. The dots $\cdots$ labeling the quantum wire bundle represent all qubits affected by the parity string $\mathcal{Z}_{pq}^{rs}$.}
    \label{fig:DoubleExcitation}
\end{figure*}

The generator of the double excitation from modes $p, q$ to $r, s$ is decomposed as
\begin{align}
    & G_{pq}^{rs} \to \frac{1}{8} \mathcal{Z}_{pq}^{rs} \label{eq:DoubleExcitationGeneratorPauli}\\
    &\times\left(
    X_pY_qY_rY_s + Y_pX_qY_rY_s - Y_pY_qX_rY_s - Y_pY_qY_rX_s
    \right. \nonumber \\
    & \left. 
    -Y_pX_qX_rX_s - X_pY_qX_rX_s + X_pX_qY_rX_s + X_pX_qX_rY_s
    \right), \nonumber
\end{align}
with the parity string $\mathcal{Z}_{pq}^{rs}\coloneqq\prod_{j\in\{p, q, r, s\}}\bigotimes_{k<j}Z_k$.
It involves two different sorts of Pauli strings, namely all permutations of $YXXX$ and $XYYY$ across the four orbitals $p,q,r,s$. 

In the following, we assume that $p < q < r < s$, thus the MS gates act on the $n=(q-p) + (s-r) +2$ qubits affected by the double excitation. In case that $q=p+1$ and $s=r+1$, the double excitation acts precisely on the 4 qubits $p,q,r,s$. These are the qubits we can generally use to deploy the $R_z$ gates. Since the generator entails 8 strings, which we have to distribute among 4 qubits, we can not implement all strings at once. Instead, we need to distribute the rotations among two different layers, amounting to a minimum of 4 MS gates.

As introduced in the \hyperref[sec:Introduction]{introduction}, 
for an even number of qubits $n=2m$, the $YXXX$-type strings can be readily realized in three layers using the \textbf{XX} gate: 
\begin{align}
    &\textbf{XX} (-Z_p - Z_q + Z_r + Z_s) \textbf{XX}^\dagger = (-1)^m \mathcal{X}_{pq}^{rs}\times
    \label{eq:DoubleExcitationGeneratorDecompositionXXXYEven} \\
    &(-Y_pX_qX_rX_s - X_pY_qX_rX_s + X_pX_qY_rX_s + X_pX_qX_rY_s) \nonumber.
\end{align}
For the $XYYY$-type strings, the same result can be achieved using \textbf{YY} interactions instead, thus circumventing the need for additional local transformations on the qubits $p, q, r, s$: 
\begin{align}
    &\textbf{YY}(-Z_p - Z_q + Z_r + Z_s) \textbf{YY}^\dagger = (-1)^m \mathcal{Y}_{pq}^{rs}  \label{eq:DoubleExcitationGeneratorDecompositionYYYXEven} \\
    &\times (X_pY_qY_rY_s + Y_pX_qY_rY_s - Y_pY_qX_rY_s - Y_pY_qY_rX_s). \nonumber
\end{align}
For an odd number of qubits $n=2m+1$, the results of equations~\ref{eq:DoubleExcitationGeneratorDecompositionXXXYEven} and \ref{eq:DoubleExcitationGeneratorDecompositionYYYXEven} can be achieved similarly to equation~\ref{eq:SingleExcitationGeneratorDecompositionOdd} by employing the identities $X=HZH$  and $Y=\sqrt{X}^\dagger Z\sqrt{X}$. 
The resulting circuits can be inferred from Fig.~\ref{fig:DoubleExcitation}.  Compared to Ref.~\cite{casanova2012quantum}, our technique reduces the number of MS gates from 16 down to 4, which is optimal.

If we relieve the constraint that excitations shall only occur from occupied to virtual orbitals, we can employ additional parallelizations. All distinct permutations of $G_{pq}^{rs}$, i.e.,~$G_{pr}^{qs}$ and $G_{ps}^{qr}$ give rise to the same eight Pauli strings, hence they can be implemented with the same cost as one double excitation by adjusting the angles in Fig.~\ref{fig:DoubleExcitation}. We stress that this observation is not unique to our circuits and has already been efficiently employed in, e.g.,~Ref.~\cite{kornell2023improvements}. 

A double excitation where two indices are identical effectively boils down to a controlled single excitation
\begin{equation}
    G_{pj}^{qj} = -ia^\dagger_j a_j (a_p^\dagger a_q - \text{H.c}) = -n_j G_p^q,
    \label{eq:ControlledSingleExcitation}
\end{equation}
where $n_j \coloneqq a_j^\dagger a_j$ is the particle number operator, which makes $j$ act as a control mode.
While these types of excitations are typically neglected in UCCSD theory, they arise in generalized UCC theory, such as UCCGSD \cite{lee2018generalized}. These terms further appear in the simulation of electronic structure Hamiltonians, which we will exploit later (cf.~\hyperref[subsec:HamiltonianSimulation]{results}). 
In addition, normal- and controlled single excitations are universal for particle-number preserving operations \cite{Arrazola2022universalquantum}. The controlled-singles circuits arise naturally from the regular singles circuits in Fig.~\ref{fig:SingleExcitation} by replacing the $R_z(\theta)$ gates by controlled rotations. We leave the technical details to the \hyperref[subsec:ControlledExcitation]{methods} section.

\subsection*{Circuit Costs for Higher Order Excitations}

The expressivity of the UCCSD ansatz can be increased by including triple excitations (UCCSDT) \cite{haidar2023extension} or even higher order terms \cite{anand2022quantum}. The generator of an $N$-th order excitation generally assumes the form of $2^{2N-1}$ mutually commuting Pauli strings under the JW mapping, where each string consists of an odd number of $X$ and $Y$ operators \cite{Romero2019Strategies}. Using the same parallelization strategy as for the singles and doubles, we can always implement subsets of $N$ strings in parallel. Therefore, our approach reduces the MS count from $2^{2N}$ down to $2\lceil 2^{2N-2}/N \rceil$, thus providing an $\mathcal{O}(N)$ speedup. 

We note that for any excitation order $N$, in principle, it is possible to capture the non-locality of the JW strings using only 2 global MS gates, at the start and end of the decomposition. The difference to our approach with multiple global MS gates then lies in the fact that the two MS gates are interspersed with all odd-parity ($1, 3,\dots, 2N-1$) non-local Pauli-$Z$ rotations, which are however local to the $2N$ qubits subject to the excitation. These could then in return again be decomposed in terms of smaller MS gates. However, the exponential cost is not removed, but rather relegated to the inner decomposition. More details on this mixed approach are provided in Supplementary Note \hyperref[app:CNOT]{4}.  

\subsection*{Applicability to other Fermion-To-Qubit Mappings}

We now shortly address if and how our parallelization techniques can be applied to other fermion-to-qubit mappings. 
For a simple illustration, we consider the generator of a single-excitation $G_1^3$ for a system with 4 fermionic modes, e.g.~the H$_2$ molecule with a minimal active space of 2 electrons and four spin-orbitals, within the Parity \cite{seeley2012bravyi} and the Bravyi-Kitaev (BK) \cite{bravyi2002fermionic} mappings. 
With the Parity mapping we obtain
\begin{align}
    G_1^3=\frac{1}{2} \left( Z_0 X_1 Y_2 - Y_1 X_2 Z_3 \right)
    \label{eq:SingleExcitationPM13_main},
\end{align}
while the BK mapping yields
\begin{align}
    G_1^3=\frac{1}{2} \left( Z_0 Y_1 Z_2 - Y_1 Z_3 \right).
\end{align}
The main insight is that for both mappings, a fermionic excitation potentially maps to Pauli strings acting on different subsets of qubits.
In both examples, the two Pauli strings have different supports, and hence cannot be parallelized through means of equations~\ref{eq:SingleExcitationGeneratorDecompositionEven} and~\ref{eq:SingleExcitationGeneratorDecompositionOdd}.
If one would like to implement the single excitation from equation~\ref{eq:SingleExcitationPM13_main} with MS gates using the decomposition technique from equation~\ref{eq:Generator}, one would need two MS gates for $Z_0 X_1 Y_2$ and another pair of MS gates for $Y_1 X_2 Z_3$. This results in more than the minimum amount of needed MS gates compared to JW. Since the Pauli strings in the parity mapping have the same $\mathcal{O}(N)$ locality scaling as in JW, we conclude that the parity mapping is less suitable for efficient simulations with global MS gates. Within the BK mapping, while the parallelization also fails, having more MS gates with only $\mathcal{O}(\log(N))$ locality might still prove beneficial, since the gates would be faster and less erroneous.

\subsection*{Applicability to Qubit Excitations} \label{subsec:QEB}
Qubit excitations have come up as a resource-friendly alternative to fermionic excitations for UCCSD theory \cite{Yordanov2021Qubit}. In principle, qubit excitations differ from fermionic excitations in that they simply ignore the parity string in the generator, thus avoiding operations on the parity qubits of the excitations, and making the operation more resource-friendly. While this does not properly capture the fermionic character of the system, it works extremely well for variational problems such as UCCSD. We want to stress that all our circuits hold equivalently for qubit-excitations when removing the parity-qubits. Thus, our qubit single-excitations entails precisely two 2-qubit MS gates, and our qubit double-excitations consist of four 4-local MS gates. Our techniques are thus readily applicable to more hardware-friendly qubit excitation-based (QEB) notion of UCCSD. 

\begin{figure*}[t]
    \centering
    \includegraphics[width=\textwidth]{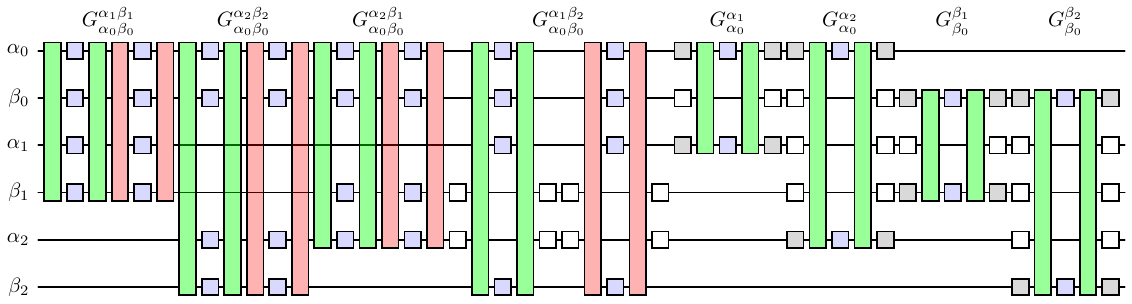}
    \caption{Schematic circuit decomposition of one layer of the UCCSD ansatz in first-order Trotterization. We use the same coloring scheme as for the previous figures; \textbf{XX} gates in green, \textbf{YY} gates in red, $R_z$ gates in blue, local Cliffords in white if they are due to the parity string, gray if they account for odd numbers of qubits in the interaction. Note that some adjacent local Clifford gates cancel out and are only explicitly depicted for the sake of clarity.}
    \label{fig:UCCSD}
\end{figure*}

\subsection*{Circuits for Hamiltonian Simulation} \label{subsec:HamiltonianSimulation}

Concerning the symmetrized terms in the Hamiltonian from equation~\ref{eq:SymmetrizedHamiltonian}, we can trace them back to the antisymmetrized structure from UCCSD by exploiting the local equivalence of the antisymmetrized electron terms (single- and double- excitations) and the symmetrized electron terms in fermionic space:
\begin{align}
     \tilde G_p^q &= \exp\left(-i\frac{\pi}{2}n_p\right) G_p^q \exp\left(i\frac{\pi}{2}n_p\right), \label{eq:LocalEquiv1}\\
     \tilde G_{pq}^{rs}
     &= \exp\left(-i\frac{\pi}{2}n_p\right) G_{pq}^{rs}\exp\left(i\frac{\pi}{2}n_p\right). \label{eq:LocalEquiv2}
\end{align}
Note that one could also use other particle number operators than $n_p$ involved in the excitation ($q$ for singles and $q,r,s$ for doubles), but then for the orbitals in the superscript we have to replace $\pi/2\to -\pi/2$.
Also, for a controlled excitation $G_{pj}^{qj}$, only the modes $p$ and $q$ can be used. 
For more details, refer to Supplementary Note \hyperref[app:Equivalence]{5}.
Under the JW mapping, this local fermionic equivalence manifests as a local Clifford equivalence, i.e.,~$\exp{(-i\pi/2n_p)} \to S_p$ (up to a global phase which cancels out with the conjugate term), where $S_p$ is the $S$-gate acting on qubit $p$. 
This way, we entirely avoid mapping out $\tilde G$ with the JW mapping and instead can recycle the circuits from Figs.~\ref{fig:SingleExcitation},~\ref{fig:DoubleExcitation}, and~\ref{fig:ControlledSingleExcitation}. We point out that, unlike in variational applications, the phase $(-1)^m$ is important here to avoid accidentally performing backwards time evolution.

The only terms that can not be traced back to the excitation circuits are the density terms $\tilde G_p^p$ and the Coulomb repulsion terms $\tilde G_{pq}^{pq}$, which under the JW transformation map to
\begin{align}
    \tilde G_p^p &= 2n_p \to I-Z_p, \quad \text{and} \label{eq:ParticleNumberPauli}\\
    \tilde G_{pq}^{pq} &= -2n_pn_q \to \frac{1}{2} (-I + Z_p + Z_q - Z_p Z_q). \label{eq:CoulombRepulsionPauli}
\end{align}
These terms are at most 2-local, and the required $R_{zz}$ gate for the $ZZ$ terms can again be decomposed into MS gates as detailed in the \hyperref[sec:Introduction]{introduction}. 

We supplement this section with the same remark as before - 
that all permutations of $G_{pq}^{rs}$ give rise to the same string pool, and can thus be fully parallelized (the same holds separately for $\tilde G$). Since for the two-electron integrals we generally have $h_{pqrs}\neq h_{prqs} \neq h_{psqr}$, the Pauli strings corresponding to $G_{pq}^{rs}$, $G_{pr}^{qs}$ and $G_{ps}^{qr}$ will in general not cancel out. 

Finally, it is worth pointing out that real-valued orbitals effectively remove the antisymmetrized terms (equation~\ref{eq:AntisymmetrizedHamiltonian}), thus cutting the number of non-local terms by half. In addition, more symmetries can be exploited, which however only reduce the $R_z$ count and not the number of MS gates. We provide the technical details in the \hyperref[app:RealOrbitalSimulation]{methods} section. 

\subsection*{Computational Goals of the Circuits}
Our circuits are applicable to both time-dependent and time-independent studies. However, the time dependency brings an additional computational difficulty due to the coupling of electronic- and nuclear states. To tackle down this obstacle, techniques were developed leading to two types of quantum dynamics \cite{Marx_Hutter_2009,Richings_2015}. Either only the electronic motion is treated with quantum mechanics and the nuclear motion is evaluated classically, or both electronic and nuclear motions are treated with quantum mechanics. In the following, these two case are referred to as mixed classical-quantum dynamics, and fully-quantum dynamics, respectively. 

The first type of dynamics, the mixed classical-quantum dynamic (MCQD) method, differs from molecular dynamics methods where the system is only treated classically and interatomic potentials are used to mimic interatomic interactions \cite{Hansson_2002}, whereas the electronic Schrödinger equation is not explicitly considered. For MCQD methods, the nuclei are also treated classically with Newton's equation of motion, and the Schrödinger equation is solved solely for the electrons \cite{Marx_Hutter_2009}. Among the MCQD methods, either the dynamic is performed within a single electronic state, or the electronic wave-function evolves onto several electronic states and non-adiabatic couplings are involved in order to transfer the wave-function from one state to another. In any case, all electronic states in a MCQD method are determined within the Born-Oppenheimer approximation.

When only a single adiabatic electronic state (usually the ground-state) is considered, the electronic wavefunction is propagated onto this single potential energy surface, and all nuclei have a classical motions. This can be applied to describe atomic interactions in ground states \cite{Iftimie_2005}, such as diffusion properties \cite{He_2018}, reaction processes \cite{Mosconi_2015}, and also the calculation of observables such as IR spectra \cite{Thomas_2013}. One can use our UCCSD circuits to prepare a single adiabatic electronic state (usually the ground-state), and then slowly evolve it in a classical-quantum hybrid loop, where the UCCSD ansatz is used to update the electronic state based on the changed nuclear geometry.

When multiple adiabatic electronic states are considered \cite{crespo-otero_recent_2018}, the electronic wavefunction can propagate among them through classical motion of the nuclei. This is formalized as the trotterized time evolution of a time-dependent electronic Hamiltonian, as detailed in Ref.~\cite{bultrini_mixed_2023}.
Among the wide range of applications of such method, one can name photoisomerization processes \cite{Wang_2016}, chemical processes in the atmosphere \cite{Curchod_2024}, and simulating the behavior of molecular aggregates for optoelectronics \cite{Hernández_2023}. For such cases, one could start with UCCSD to prepare the ground state, and then non-adiabatically evolve it in time using the circuits for Trotterized Hamiltonian simulation. 

The second type of dynamics entails a fully quantum mechanical treatment of both electrons and nuclei, and the Born-Oppenheimer approximation is not considered. 
These methods are very attractive for the time-study of quantum systems in the realm of non-adiabatic processes among several electronic states \cite{Nelson_2020}. They are well known for studying internal conversion in conjugated chromophores (chemical processes involving bond breaking \cite{Nelson_2016}, carbon nanotube\cite{Wong_Lee_2011}, non-radiative relaxation in chlorophylls\cite{Fidler_2014}), energy transfer in molecular aggregates (dendrimers \cite{Fernandez_2009}, or vibrational studies in OLEDs\cite{Nelson_2017}). By encoding the vibrational modes of the nuclear motion into the normal modes of the ion chain of an ion trap computer \cite{Olaya_2025}, one could employ our time evolution circuits in a digital-analog fashion, as detailed in \cite{casanova2012quantum, lamata2014efficient}. A state-of-the-art implementation of such procedure is provided in Ref.~\cite{Ha_MacDonell_2025}.

\begin{figure*}[t]
    \centering
    \includegraphics[width=\textwidth]{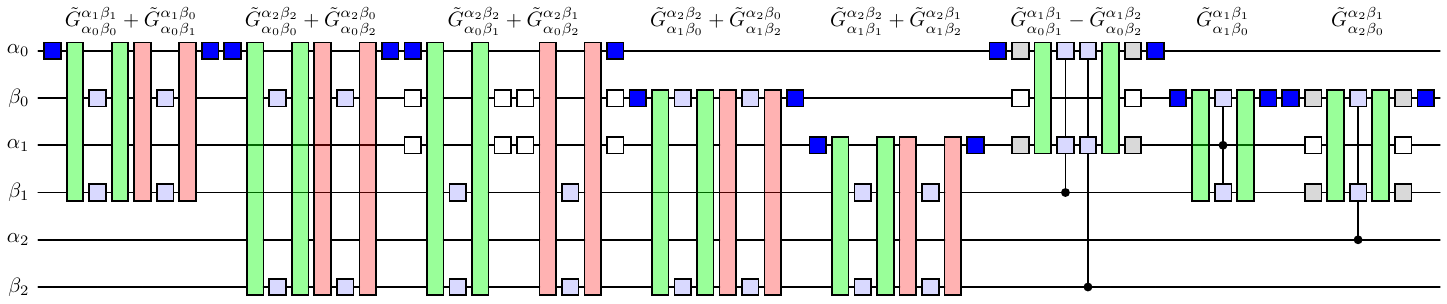}
    \caption{Schematic circuit decomposition of one Trotter step of $\exp(-i\delta t H_\text{non-loc.})$ in first-order Trotterization. We use the same coloring scheme as for the previous figures; \textbf{XX} gates in green, \textbf{YY} gates in red, $R_z$ gates in blue, local Cliffords in white if they are due to the parity string, gray if they account for odd numbers of qubits in the interaction. In addition, we introduce dark blue gates representing the $S^{(\dagger)}$ gates ensuring the symmetrization. Note that some adjacent local Clifford gates cancel out and are only explicitly depicted for the sake of clarity.}
    \label{fig:TrotterStep}
\end{figure*}

\subsection*{Example 1: A UCCSD Layer}

We now provide some illustrative examples on how to assemble practical fermionic circuits within our gate decompositions. 

Here, we demonstrate our technique on the \ce{H_3^+} molecule in the \texttt{STO-3G} basis set. This system entails 2 electrons distributed among 6 spin-orbitals, and thus provides a minimalist example with non-localities arising from the JW mapping in both the single- and double-excitations (or quadratic and quartic Hamiltonian terms). Note that we alternate the spin-up $(\alpha)$ and spin-down $(\beta)$ orbitals in our state and operator notation, i.e.,~$\ket{\alpha_0,\beta_0, \alpha_1, \beta_1, \alpha_2, \beta_2}$. We use the same order to enumerate the orbitals in the JW mapping.

We start off by constructing the circuit for one first-order Trotter step in UCCSD theory. For that purpose, we are not concerned with the precise structure of the Hamiltonian, but rather the Hartree-Fock ground state $\ket{\psi}_\text{HF} = \ket{110000}$ and the eligible excitations starting from that state. Here, there are 4 unique spin-preserving single excitations
$G_{\alpha_0}^{\alpha_1}, G_{\alpha_0}^{\alpha_2}, G_{\beta_0}^{\beta_1}$, 
and 
$G_{\beta_0}^{\beta_2}$.
In addition, there exist 4 different spin-preserving double excitations 
$G_{\alpha_0, \beta_0}^{\alpha_1, \beta_1}, G_{\alpha_0, \beta_0}^{\alpha_2, \beta_1}, G_{\alpha_0, \beta_0}^{\alpha_1, \beta_2}$ 
and
$G_{\alpha_0, \beta_0}^{\alpha_2, \beta_2}$. 
The quantum circuit corresponding to these excitations is schematically depicted in Fig.~\ref{fig:UCCSD}. Our circuit employs 24 MS gates while a string-by-string implementation amounts to 80 gates, hence we achieve a gate reduction by a factor of $\sim 3.3$.

\subsection*{Example 2: A Trotter Step of the Hamiltonian}

We compute the one- and two-electron integrals with the  \texttt{STO-3G} basis set in the equilibrium geometry, i.e., a bond distance of \SI{0.784}{\angstrom} and a bond angle of \SI{60}{\degree} using the \texttt{pyscf} package \cite{sun2018pyscf, sun2020recent}. This gives rise to the Hamiltonian $H=H_\text{loc.} + H_\text{non-loc.}$ (we list the most significant terms in units of \SI{1}{Ha}), where the local part containing terms of the type $\tilde G_p^p$ and $\tilde G_{pq}^{pq}$ is given by

\begin{align}
    H_\text{loc.} 
    &= -0.917 (\tilde G_{\alpha_0}^{\alpha_0} + \tilde G_{\beta_0}^{\beta_0}) \nonumber \\
    &\hphantom{=~} - 0.535 (\tilde G_{\alpha_1}^{\alpha_1} + \tilde G_{\alpha_2}^{\alpha_2} + \tilde G_{\beta_1}^{\beta_1} + \tilde G_{\beta_2}^{\beta_2}) \nonumber \\
    &\hphantom{=~} - 0.337 (\tilde G_{\alpha_1\beta_1}^{\alpha_1\beta_1} + \tilde G_{\alpha_2\beta_2}^{\alpha_2\beta_2}) - 0.307 \tilde G_{\alpha_0\beta_0}^{\alpha_0\beta_0} \nonumber \\
    &\hphantom{=~} -0.298 (\tilde G_{\alpha_0\beta_2}^{\alpha_0\beta_2} + \tilde G_{\alpha_2\beta_0}^{\alpha_2\beta_0} + \tilde G_{\alpha_0\beta_1}^{\alpha_0\beta_1} + \tilde G_{\alpha_1\beta_0}^{\alpha_1\beta_0}) \nonumber \\
    &\hphantom{=~} -0.265 (\tilde G_{\alpha_1\beta_2}^{\alpha_1\beta_2}+\tilde G_{\alpha_2\beta_1}^{\alpha_2\beta_1}) \nonumber \\
    &\hphantom{=~} -0.229 (\tilde G_{\alpha_1\alpha_2}^{\alpha_1\alpha_2} + \tilde G_{\beta_1\beta_2}^{\beta_1\beta_2}) \label{eq:H3+HamiltonianLocal} \\
    &\hphantom{=~} -0.226 (
    \tilde G_{\alpha_0\alpha_1}^{\alpha_0\alpha_1} + \tilde G_{\alpha_0\alpha_2}^{\alpha_0\alpha_2} + \tilde G_{\beta_0\beta_1}^{\beta_0\beta_1} + \tilde G_{\beta_0\beta_2}^{\beta_0\beta_2}), \nonumber 
\end{align}
while the non-local part reads 
\begin{align}
    H_\text{non-loc.} 
    &= -0.142 (\tilde G_{\alpha_0\beta_0}^{\alpha_1\beta_1}  + \tilde G_{\alpha_0\beta_1}^{\alpha_1\beta_0} + \tilde G_{\alpha_0\beta_0}^{\alpha_2\beta_2}+ \tilde G_{\alpha_0\beta_2}^{\alpha_2\beta_0}) \nonumber \\
    &\hphantom{=~} -0.090 (\tilde G_{\alpha_0\beta_1}^{\alpha_1\beta_1} + \tilde G_{\alpha_1\beta_0}^{\alpha_1\beta_1} - \tilde G_{\alpha_0\beta_2}^{\alpha_1\beta_2} - \tilde G_{\alpha_2\beta_0}^{\alpha_2\beta_1}) \nonumber \\
    &\hphantom{=~} +0.090 (\tilde G_{\alpha_0\beta_1}^{\alpha_2\beta_2} + \tilde G_{\alpha_0\beta_2}^{\alpha_2\beta_1} + \tilde G_{\alpha_1\beta_0}^{\alpha_2\beta_2}+ \tilde G_{\alpha_1\beta_2}^{\alpha_2\beta_0}) \nonumber \\
    &\hphantom{=~} -0.072 (\tilde G_{\alpha_1\beta_1}^{\alpha_2\beta_2} + \tilde G_{\alpha_1\beta_2}^{\alpha_2\beta_1}). \label{eq:H3+HamiltonianNonLocal}
\end{align}
Note that $\exp(-i\delta tH_\text{loc})$ trivially boils down to $R_z$ and $R_{zz}$ rotations according to equations~\ref{eq:ParticleNumberPauli} and \ref{eq:CoulombRepulsionPauli}. For that reason, we focus on the circuit decomposition concerning the non-local interactions. 

Due to the symmetries for real basis sets, every term $G_{pq}^{rs}$ with $p\neq q\neq r\neq s$ is accompanied by a term $G_{ps}^{rq}$ which can be included without any additional MS gates. We can further exploit that the terms $\tilde G_{\alpha_0\beta_1}^{\alpha_1\beta_1}$ and $\tilde G_{\alpha_0\beta_2}^{\alpha_1\beta_2}$ correspond to the same excitation controlled by different modes, and can thus be parallelized as well. Last, we want to emphasize the controlled excitation $\tilde G_{\alpha_1\beta_0}^{\alpha_1\beta_2}$. Here, $\alpha_1$ is the control but simultaneously part of the JW string $\mathcal{Z}_{\beta_0}^{\beta_1}$. 
Using all these properties allows us to implement $\exp(-i\delta t H_\text{non-loc.})$ with a total of 8 building blocks based on Figs.~\ref{fig:DoubleExcitation} and \ref{fig:ControlledSingleExcitation}, as we depict in Fig.~\ref{fig:TrotterStep}. Our circuit entails 26 MS gates whereas the implementation of each string separately (also using the symmetries and controlled rotations for the sake of comparability) amounts to 56, enabling a speedup of $\sim 2.2$. A naive implementation of each excitation separately without the use of symmetries gives rise to 176 MS gates, showcasing that the main benefit here stems from the symmetry exploitation rather than the parallelization.

\subsection*{Noise Modeling} \label{sec:noise_modelling}

So far, we have assumed ideal maximally entangling MS gates in our circuit decompositions. In practice on noisy quantum simulators, shorter circuits do not always guarantee results with higher fidelity. Our scenario is arguably more specific, given that we carry out the same types of gates on the same qubits as in Ref.~\cite{casanova2012quantum}, just fewer times. Naively, one could estimate that we reduce the noise-strength for single,- and double excitations by factors of 2, and 4, respectively. Such simple scaling behavior holds, e.g.~for depolarizing channels, but is not accurate for realistic noise models and experiments \cite{Koenig2024ZNE}. Therefore, the purpose of this section is to develop a realistic ion trap noise model, which we 
employ to assess if our reductions in MS gates actually result in higher fidelity building blocks for fermionic circuits.

We consider 12 \ce{^171Yb^+} ions trapped in a linear harmonic radio-frequency Paul trap \cite{Deslauriers2004Paul}, with axial trap frequency $\omega_z = 2\pi \times \SI{0.5}{\mega \hertz}$ and radial trap frequency $\omega_x = 2\pi \times  \SI{3.33}{\mega\hertz}$, such that the distances between adjacent ions lie in the range of \SIrange{2.22}{3.21}{\micro\meter}.
We assume that each ion in the chain can be individually addressed with its own pair of red- and blue side-band Raman beams with optical frequencies $\omega_0 \pm \mu_i(t)$, where $\mu_i(t)$ is the time-dependent Raman detuning with respect to the spin resonance frequency $\omega_0 = 2\pi \times \SI{12.643}{\giga\hertz}$ \cite{Olmschenk2007Manipulation} between the two spin-$1/2$ states which serve as the qubit computational basis states. 
We further assume that both Raman beams have equal time-dependent pulse amplitudes $\Omega_i^{(1)}(t)=\Omega_i^{(2)}(t)$ and originate from the same pulse-modulated laser source. Thus, AC stark shifts due to laser power fluctuations may be neglected as the contribution from the red and blue shifted Raman beams cancel each other out \cite{Morong2023Engineering}. 
Under the rotating approximation $\omega_0 \gg \mu(t)$, and the Lamb-Dicke limit, the Hamiltonian of the experiment may be written as \cite{Morong2023Engineering, blumel2021power}
\begin{align}
    \label{eq:MSHamiltonianExp}
    H_\textrm{MS}(t) &= \sum_i \sum_p  \eta_{p}^{(i)} g^{(i)}(t) \left[a_p e^{-i\omega_p t} + a_p^\dagger e^{i\omega_p t}\right]  \\
    &\hphantom{=~} \times \left[\cos(\phi)X_i + \sin(\phi) Y_i\right] \nonumber,
\end{align}
where the indices $i$ and $p$ label the qubits/ions and radial motional modes, respectively.  The Lamb-Dicke parameters $\eta_{p}^{(i)} \propto b_{p}^{(i)} /\sqrt{\omega_p}$ describe the coupling strength between qubit $i$ and radial mode $p$ and entail the normal mode transformation matrix $b_{p}^{(i)}$ of ion $i$ and motional mode $p$ with frequency $\omega_p$, and $a_p$ and $a_p^\dagger$ are the normal mode phonon creation- and annihilation operators. The $g^{(i)}(t) \coloneqq \Omega^{(i)}(t) \sin\left(\int_0^t d\tau \mu^{(i)}(\tau)\right)$ are the pulse functions defined through the aforementioned amplitudes and detunings.
The phase $\phi$ is determined by the phase of the two Raman beams and is chosen such that either $\phi=0$ for the \textbf{XX}-, or $\phi=\pi/2$ for the \textbf{YY}-interaction. The full experimental MS gate unitary corresponding to the Hamiltonian from equation \ref{eq:MSHamiltonianExp} is described in the \hyperref[subsec:MSGateEXP]{methods} section. 
\\

For the modeling of different experimental error sources, we mostly follow Ref \cite{Lotshaw2023Modeling}, which introduces a noise model for the global MS gate showing good agreement with experimental data. In particular, the authors consider vibrational mode frequency fluctuations $\omega_{p} \to \omega_{p} + \Delta \omega_{p}$, which lead to instabilities in both the geometric phases and displacements. Assuming thermal initial states of the vibrations \cite{Roos2008Ion}, the decoherence from residual entanglement between electronic and vibrational states is further amplified. Laser power fluctuations of the two laser beams driving the MS interaction lead to fluctuations in the Rabi rates $\Omega^{(i)}(t) \to \Omega^{(i)}(t)(1+\Delta \Omega^{(i)})$ \cite{blumel2021power, Lotshaw2023Modeling}. In addition, state preparation and measurement (SPAM) errors are considered.
In their experiments, the authors find that vibrational frequency fluctuations and SPAM errors are the dominant sources of error. 
We focus on the former, given that scalable readout error mitigation methods exist \cite{Nation2021Scalable}, which could in principle handle the fundamental detection infidelity limit of \ce{^171Yb^+} ions which is about $10^{-3}$ \cite{Noek2013Fast}. 
We therefore consider a pulse modulation scheme which is robust against vibrational frequency fluctuations \cite{blumel2021power}. We provide a qualitative description in the \hyperref[subsec:PulseModulation]{methods} section, 
whereas an in-depth guide is provided in Supplementary Note \hyperref[app:LinearQuadraticConstraints]{7}. For solving the optimization problem outlined in Sec.~\ref{subsec:PulseModulation}, we use PyOptInferface \cite{Yang2024PyOptInterface, Yang2025Accelerating} in combination with the Ipopt \cite{Waechter2005On} software.
\\

Assuming frequency fluctuations of $\SI{1}{\kilo\hertz}$ \cite{blumel2021power} and power fluctuations of $\SI{0.5}{\%}$, we find a two-qubit gate fidelity of $\mathcal{F}_2=\SI{99.97}{\%}$, and observe an exponential decay in fidelity when increasing the number of qubits, up to a fidelity of $\mathcal{F}_{12}=\SI{98.27}{\%}$. We find that the decay is well described by $\mathcal{F}_2^{N(N-1)/2}$, which is motivated by the fact that an $N$-qubit MS gate implicitly entails $N(N-1)/2$ two-qubit MS gates. Another important feature is that the gate time of our optimized MS gates scales sub-linearly in the number of qubits, which conveniently yields sub-linear circuit times despite the linear locality of the JW mapping. A full characterization of the noise model is provided in Supplementary Note~\hyperref[app:NoiseModel]{8}. More specifically, the assumed Lamb-Dicke parameters are shown in \hyperref[tab:MotionalModes]{Supplementary Table 1}, the optimized pulses are displayed in \hyperref[fig:pulses]{Supplementary Figure 1}, the gate fidelities as a function of the number of qubits are provided in \hyperref[fig:MSGATE_BENCHMARK]{Supplementary Figure 2}, and last the average laser power as a function of the number of qubits and gate time is analyzed in \hyperref[fig:PULSETIME]{Supplementary Figure 3}.

\begin{figure*}[t]
    \centering
    \includegraphics[width=0.8\textwidth]{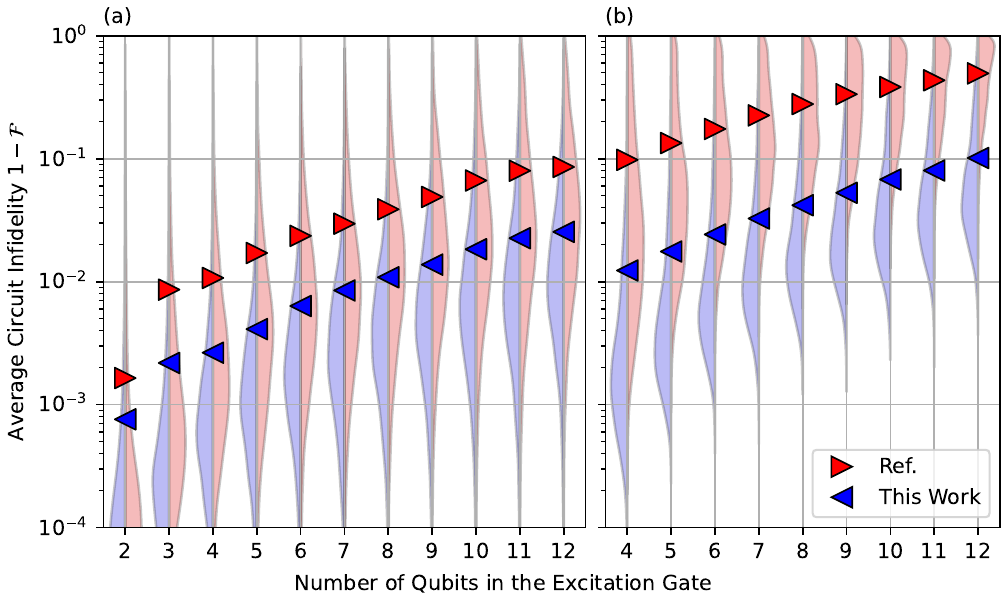}
    \caption{Circuit infidelities for (a) single excitations, and (b) double excitations, for different number of qubits the gates act on, once with our decompositions (blue), and once with the one presented in Ref.~\cite{casanova2012quantum} (red). The statistics come from first sampling the frequency noise parameters paired with random excitation angles $M=\num{5000}$ different times for each number of qubits. We then consider laser power fluctuations for each MS gate separately, thus resulting in averages over $\num{10000}$, and $\num{20000}$ experimental realizations of MS gates in our circuits for singles-, and doubles, respectively. Consequently, for the reference, we consider $\num{20000}$, and $\num{80000}$ realizations of MS gates. The markers show the mean of the respective distributions.}
    \label{fig:benchmark_exc_gates}
\end{figure*}

\subsection*{Excitation Circuit Benchmarks} \label{sec:benchmarks}

We first benchmark individual excitation gate decompositions, and, second, benchmark approximate ground state preparation of various molecules.
We assume that single-qubit gate errors would be negligible and only apply noise to the MS gates according to the \hyperref[sec:noise_modelling]{results} section. 
We keep the frequency fluctuations $\Delta \omega_p$ fixed for every MS gate during one circuit execution, but assume different laser power fluctuations $\Delta \Omega^{(i)}$ for each MS gate within that circuit. This is motivated by the fact that the frequency fluctuations can be assumed constant on the timescale of minutes \cite{blumel2021power}. For all following simulations, we approximate the infinite sampling limit by computing expectation values based on statevector simulations for $M$ noisy realizations of the circuits using Qiskit \cite{JavadiAbhari2024Quantum}. 

For benchmarking the individual excitation gates, we evaluate the average circuit infidelity $1-\mathcal{F}$ with 
\begin{equation}
    \mathcal F = \frac{1}{2\pi} \int_{-\pi}^\pi d\theta \left|\bra{\psi_\mathrm{HF}}U^\dagger(\theta)\tilde U(\theta)\ket{\psi_\mathrm{HF}}\right|^2\,,
\end{equation}
for single and double excitations. Here, $\tilde U(\theta)$ is a noisy excitation gate and $U(\theta)$ is an ideal excitation gate without noise. $\ket{\psi_{\mathrm{HF}}}$ is the state where the qubits $p$, for $U_p^q$, or $p$ and $q$, for $U_{pq}^{rs}$, are in state $\ket{1}$ while all others are in state $\ket{0}$. We benchmark excitation gates that act on different number of qubits, from two to twelve, on arbitrary subsets of the linear ion trap. The results of our simulations are shown in Fig.~\ref{fig:benchmark_exc_gates}. The statistics in Fig.~\ref{fig:benchmark_exc_gates} come from sampling the frequency noise parameters ($\Delta \omega_p$) $M=5000$ different times for each locality, and then applying laser power fluctuations $\Delta \Omega^{(i)}$ to each MS gate separately. We see about half-an-order of magnitude improvements in the circuit infidelity for single excitations, when using our decomposition compared to the decomposition of Ref.~\cite{casanova2012quantum}. For double excitations, this improvement increases to about one order of magnitude. These relative improvements are rather independent of the number of qubits the excitation gates act on, yet slowly decrease with increasing number of qubits. The stronger improvement for double excitations is naturally attributed to that fact that our decomposition reduces the number of MS gates by a factor of 4 instead of 2 for single excitations.

\subsection*{UCCSD Benchmarks}

Further, we benchmark 11 molecules (\ce{H2}, \ce{HeH^+}, \ce{H3^+}, \ce{H4}, \ce{He2}, \ce{OH^-}, \ce{HF}, \ce{NeH^+}, \ce{LiH}, \ce{BeH_2}, and \ce{H2O}), which geometries and electronic structure Hamiltonians are taken from the Pennylane dataset \cite{Utkarsh2023Chemistry}. The Hartree-Fock ground state of diatomic He$_2$ has been calculated in the \texttt{6-31G} basis set, while for all other molecules, the basis set \texttt{STO-3G} was used. We apply the frozen core approximation 
to the molecules \ce{H2O} and \ce{BeH2}, in each case freezing two electrons associated to the spatial orbital with the lowest single-particle energy.

In order to apply our noise model, we first run a noiseless statevector calculation with the UCCSD ansatz optimized with ExcitationSolve \cite{jaeger2024fast} to find the optimal parameters for each molecule. We then build a new ansatz consisting of only excitations present in the UCCSD that have optimal parameter values with absolute values above a threshold of $10^{-2}$. 
We note that in practice, one would pick the most important excitations either classically, e.g.,~based on second order Møller–Plesset perturbation theory (MP2) for double excitations \cite{Romero2019Strategies}, or on a quantum device using notions of ADAPT-VQE \cite{grimsley2019adaptive} specifically tailored for arbitrary excitations \cite{jaeger2024fast}. The combination of classical selection and warm-start strategies and adaptive quantum algorithms enables not only shallow circuits, but also significant speed-ups in convergence \cite{Haas2026Efficient}. However, given that the goal of this section is to assess the quality of shallow UCCSD circuits under noise, the selection strategy to obtain the shallow circuit in the first place is secondary, and our brute-force selection based on a full UCCSD calculation is sufficient.
The UCCSD ansatz is build once with fermionic excitations, and once with qubit excitations \cite{Yordanov2021Qubit}. We use these parameter-truncated ansätze to perform our noisy simulations.  We have verified that for both fermionic- and qubit excitations, the truncated ansätze can achieve chemical accuracy of $\SI{e-3}{Ha}$ \cite{peruzzo2014variational} on an ideal device.

\begin{figure*}[t]
    \centering
    \includegraphics[width=0.95\textwidth]{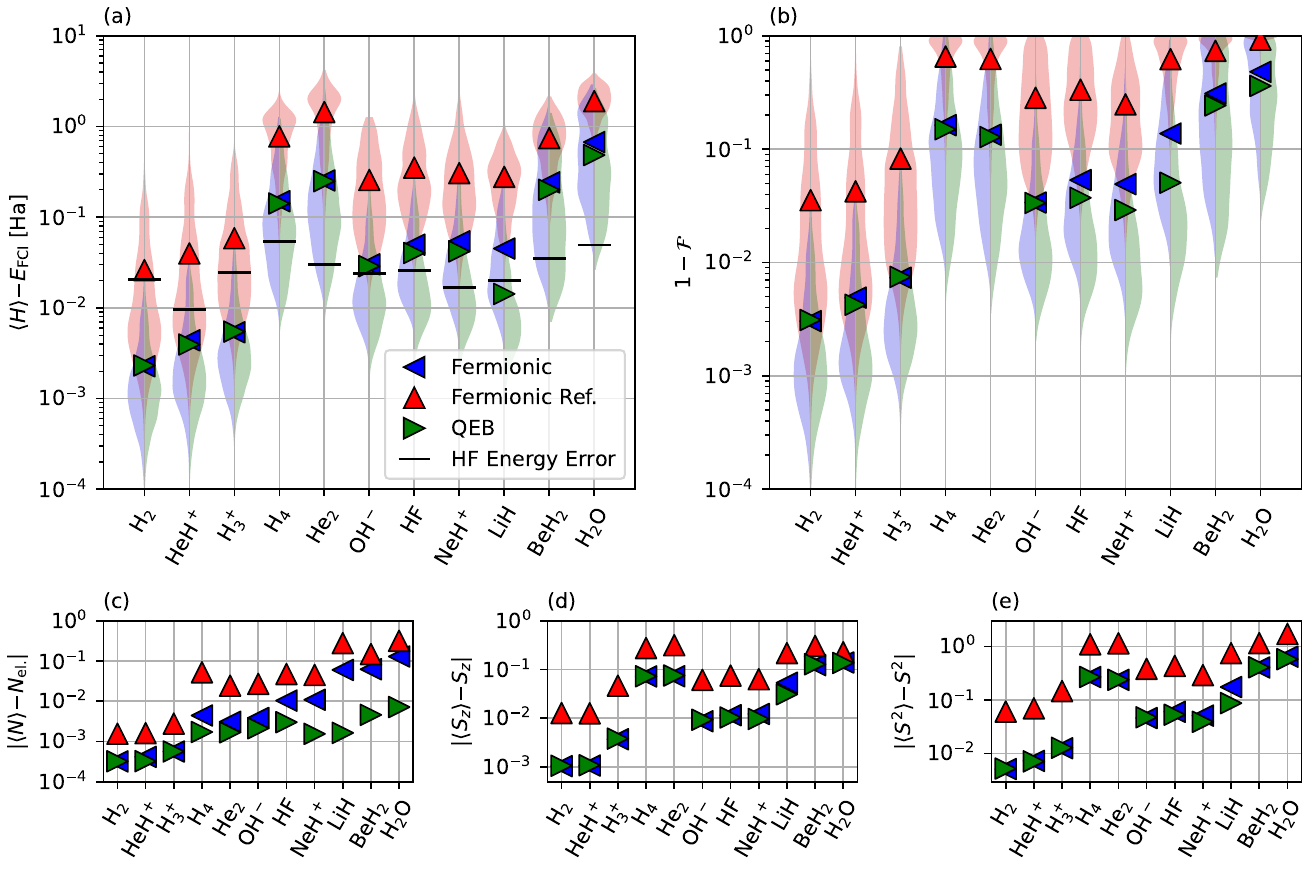}
    \caption{Results from the molecular benchmarks. (a) Energy differences to the FCI result, (b) state infidelities w.r.t. to noiseless parameter-truncated states, (c) absolute errors in the particle number operator, (d) absolute errors in the $S_z$ spin projection, and (e) absolute errors in the total spin $S^2$ for various molecules. The statistics are gathered by sampling the frequency noise parameters $M=\num{10000}$ times for \ce{H2}, \ce{HeH^+}, \ce{H3^+}, then $M=2500$ times for \ce{H4}, and $M=1000$ times for \ce{He2}, \ce{OH^-}, \ce{HF}, \ce{NeH^+}, \ce{LiH}, \ce{BeH_2}, and \ce{H2O}. Laser power fluctuations are then applied individually to each MS gate within each circuit execution. The markers show the mean of the respective distributions. We use our decompositions of fermionic excitations (blue) the decomposition from Ref.~\cite{casanova2012quantum} (red). In addition, we use our decomposition in combination with qubit excitations (green) instead of fermionic excitations. For \ce{H2O} and \ce{BeH2} we use the frozen core approximation to reduce the needed qubits by two. 
    The black horizontal lines show Hartree-Fock energy errors with respect to the FCI solution.} 
    \label{fig:UCCSD_BENCHMARK}
\end{figure*}

In Fig.~\ref{fig:UCCSD_BENCHMARK} we present energy differences to the FCI result, state infidelities w.r.t.~to noiseless parameter-truncated states, absolute errors in the particle number operator, absolute errors in the $S_z$ spin projection, and absolute errors in the total spin $S^2$. This data is reported for the various molecules listed above. The statistics in Fig.~\ref{fig:UCCSD_BENCHMARK} come from sampling the frequency noise parameters between $M=\num{10000}$ for the smallest molecular systems, down to $M=1000$ times for the largest molecules, and then applying laser power fluctuations to each MS gate within the UCCSD circuits. We perform our noisy simulations once with our decomposition of fermionic- and qubit excitations, and once with the decomposition presented in Ref.~\cite{casanova2012quantum}.

In terms of energy error, we see consistent improvements between about half-an-order and one order of magnitude for all molecules when using our decomposition. For the state infidelities we outperform the decomposition of Ref.~\cite{casanova2012quantum} by up to one order of magnitude as well. Concerning the violation of conserved quantities, we also observe significant improvements for all tested molecules.
All energy errors when using our fermionic decomposition, except for the molecules \ce{H2}, \ce{HeH^+}, and \ce{H3^+}, are still above the Hartree-Fock energy errors. This shows that the simulated noise significantly degrades the state even when only considering the most important excitations within our optimized circuit decompositions. This effect of the noise is less pronounced for the QEB circuits, where the energy error of the 12-qubit system \ce{LiH} is below the Hartree-Fock error even under noise. In general, we observe mostly small but consistent improvements for qubit excitations over fermionic excitations across all larger systems. This is particularly pronounced on the electron number error which is consistently closer to zero, which we attribute to the QEB disturbing occupation numbers on less qubits in general. On the other hand, there is only little improvement for the spin projection error and total spin error when using QEB circuits, revealing that the occupation numbers within the individual spin-sectors are more prone to errors, which in the total electron number mostly cancel out. Overall, the advantage of using QEB circuits is less pronounced than on limited-connectivity hardware, since here the number of entangling gates and circuit depth remains identical, although the locality of the MS gates is reduced depending on the type of excitation. For the considered molecules, the list of important excitations is typically dominated by doubles acting on two spatial orbitals $p, q$, i.e.~$G_{\alpha_p\beta_p}^{\alpha_q\beta_q}$, where fermionic- and qubit excitations turn out to be equivalent when the spin-orbitals are enumerated in alternating order ($\alpha_0,\beta_0, \alpha_1,\beta_1, \dots$). In fact, for \ce{H2} and \ce{H3+}, our fermionic- and QEB circuits are identical. The advantage of qubit excitations becomes most pronounced on the larger molecules \ce{HF}, \ce{NeH^+}, \ce{LiH}, \ce{BeH_2}, and \ce{H2O}, where long-range single-excitations and doubles involving more than two spatial orbitals play significant roles. The largest benefit can be seen for \ce{LiH}, which is due to the long-range fermionic double excitations $\alpha_1,\beta_1, \alpha_2,\beta_5$, and $\alpha_1,\beta_1, \beta_2,\alpha_5$, now acting on 4 instead of 10 and 8 qubits, respectively.

Overall, we conclude that our techniques provide significant improvements over the reference with respect to every metric. We also want to stress that our circuits could be readily paired with error mitigation techniques, likely leading accurate estimates of energy errors below the HF error. Since the focus of this work is on circuit decomposition techniques and the corresponding circuit quality improvements, we have deliberately refrained from using such techniques in order to isolate the influence of our methods.

\section*{Discussion}

In this work, we introduced a parallelization scheme to reduce the number of MS gates for the implementation of fermionic excitations. Compared to previous works \cite{casanova2012quantum, lamata2014efficient}, we achieve a speedup of 2, and 4, for single- and double excitations, respectively. We further generalized our parallelization strategy to qubit single- and double excitations. We then studied our circuits in the presence of realistic experimental noise by emulating a 12-qubit linear ion trap at the pulse level, showcasing significant improvements over \cite{casanova2012quantum} across various molecular benchmarks, leading to fidelity improvements of up to one order of magnitude. We do however note that even with our improvements, meaningful accuracy improvements beyond HF theory would be tricky to achieve with the noise levels of current state-of-the-art quantum computers.
Still, looking at the relative improvements in fidelity, our circuits hold the prospect of significantly improving fermionic simulations on real ion trap devices. 
And for near-term quantum hardware, pairing our technique with QEB ansätze is a promising avenue. 

We note that our techniques could be readily extended to the simulation of fermion-boson interactions in a digital-analog fashion by encoding the bosonic operators into the vibrational modes of the ion chain \cite{Leibfried2003Quantum}, as it has been suggested in Refs.~\cite{casanova2012quantum, lamata2014efficient}. This gives access to systems such as the Fröhlich model \cite{mahan2013many}, which captures the properties of polarons  in some crystal structures \cite{franchini_polarons_2021,aihemaiti_perspective_2024}. Such simulations would extend the utility of our method beyond purely fermionic static- and dynamic properties to e.g., more complex phenomena in photochemistry \cite{bauer_quantum_2020,ollitrault_molecular_2021,ollitrault_nonadiabatic_2020}, or the study of non-adiabatic electron-nuclear dynamics \cite{Ha_MacDonell_2025}.

As we have pointed out, the parallelization techniques presented in this work are not readily applicable to other fermionic mappings. Meanwhile, the development of more efficient hardware-agnostic mappings is a flourishing field of ongoing research \cite{Miller2023Bonsai, miller2024treespilation}. However, such mappings are typically developed with limited qubit connectivity in mind, and more importantly decompositions in terms of two-qubit gates. This raises the questions whether similar endeavors
could lead to more efficient parallelizable mappings within the targeted MS gate set. 

Last, we want to emphasize that our pulse simulations suggest that the MS gate time scales sub-linearly in the number of qubits, thus effectively achieving sub-linear circuit times even within linear-locality mapping such as JW, which ultimately highlights that trapped ions are promising platform for fermionic simulation.   

\section*{Methods}

\subsection*{The Backward MS Gate} \label{app:BackwardMS}

One can entirely avoid Backward MS Gates by exploiting that MS Gates are equivalent to their \enquote{forward} counterparts \cite{Müller2011Simulating}. Here, we assume an MS interaction acting on $n$ qubits.  
\begin{equation}
    U_\text{MS}(-\theta, \phi) = 
    \begin{cases}
        U_\text{MS}(\pi - \theta, \phi), & \text{for $n$ odd}, \\
        U_\text{MS}(\pi - \theta, \phi)\bigotimes_{i}\sigma_i(\phi), & \text{for $n$ even}.
    \end{cases}
\end{equation}
For even $n$, this equivalence only holds up to local unitaries of the form $\sigma_i(\phi)\coloneqq \cos(\phi)X_i+\sin(\phi)Y_i$. 
For the fully entangling MS gates \textbf{XX} and \textbf{YY} from equations~\ref{eq:MSGateXX} and \ref{eq:MSGateYY}, this boils down to a self-inverse property $U^\dagger = U$ up to local Paulis for an odd number of qubits. More preciseliy, we have
\begin{align}
    \textbf{XX}^\dagger &= \textbf{XX} 
    \begin{cases}
        I, & \text{for $n$ odd}, \\
        \bigotimes_{i} X_i, & \text{for $n$ even},
    \end{cases}
\end{align}
and
\begin{align}
    \textbf{YY}^\dagger &= \textbf{YY} 
    \begin{cases}
        I, & \text{for $n$ odd}, \\
        \bigotimes_{i} Y_i, & \text{for $n$ even}.
    \end{cases}
\end{align}

\subsection*{The Jordan-Wigner Mapping} \label{sec:JW_map}

In order to emulate fermionic systems on a quantum computer, one needs a mapping between fermionic operators and qubit operators, i.e.,~a representation of fermionic operators in the Pauli basis. In this work, we focus on the Jordan-Wigner (JW) mapping. While often employed due to its inherent simplicity, the JW mapping faces the drawback of mapping fermionic operators on an $n$-mode system to terms with linear locality $\mathcal{O}(n)$. 
However, the linear locality of the JW mapping is to be seen as less problematic for ion trap quantum computers due to the all-to-all connectivity and native global interactions.

The fermionic (creation) annihilation operators $a^{(\dagger)}$ satisfy the canonical commutation relations $\{a_p, a_q^\dagger\}=\delta_{pq}$ and $\{a_p, a_q\} = \{a_p^\dagger, a_q^\dagger\}=0$. Under the JW mapping, these fermionic operators take the following form:
\begin{equation}
    a_p^{(\dagger)} \to \frac{1}{2} \Bigl(\bigotimes_{k < p}Z_k\Bigr) \otimes (X_p \varpm iY_p).
\end{equation}
Note that the non-locality arises from the $\mathcal{O}(n)$-local parity strings consisting of Pauli-$Z$ operators, which ensure the proper anticommutation relations. Through the course of this work, we use the MS gate to efficiently take these contributions into account. 

\subsection*{Circuits for Controlled Single Excitations} \label{subsec:ControlledExcitation}

\begin{figure*}[t]
    \centering
    \includegraphics[width=0.6\textwidth]{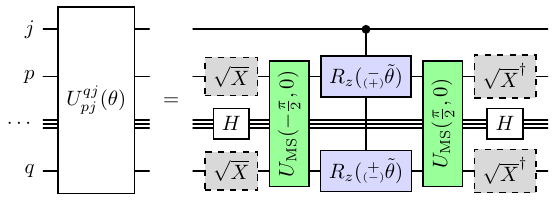}
    \caption{Circuit decomposition of the controlled single-excitation gate $U_{pj}^{qj}(\theta) = \exp(-i\theta G_{pj}^{qj})$ using the \textbf{XX} gate in terms of two MS gates and two $\texttt{C}_j R_z$ gates. 
    The dots $\cdots$ labeling the quantum wire bundle represent all qubits affected by the parity string $\mathcal{Z}_p^q$, except $j$ if $p < j < q$. The light-gray dashed gates are used if the number of qubits in the wire bundle is odd. 
    The signs in brackets account for the sign flip for $p<j<q$.}
    \label{fig:ControlledSingleExcitation}
\end{figure*}

When considering the JW-mapped expression for the controlled single excitation 
\begin{equation}
    G_{pj}^{qj} \to -\frac{1}{4} (I_j - Z_j)\mathcal{Z}_p^q \left(Y_p X_q - X_p  Y_q\right),
    \label{eq:ControlledSingleExcitationPauli}
\end{equation}
where we again assumed $p<q$, we must distinguish between two cases.\\
\textit{Case 1:}
If $j < p$ or $j > q$, the control qubit $j$ is not affected by the single-excitation generator $G_p^q$. Therefore, we can simply obtain the circuit by replacing the $R_z(\theta)$ gates in Fig.~\ref{fig:SingleExcitation} by controlled $Z$-rotations $\texttt{C}_jR_z(-\theta)$, as depicted in Fig.~\ref{fig:ControlledSingleExcitation}. Alternatively, one may implement $G_p^q$ and $Z_jG_p^q$ separately. The latter can be achieved by adding $j$ to the MS interaction. \\
\textit{Case 2:}
 If $p < j < q$, equation~\ref{eq:ControlledSingleExcitationPauli} contains the expression $-(I_j-Z_j)Z_j = I_j - Z_j$. When using controlled rotations, this effectively removes qubit $j$ from the MS interaction in Fig.~\ref{fig:SingleExcitation} and turns it into a control qubit (Fig.~\ref{fig:ControlledSingleExcitation}(a)). Optionally, the separate decomposition of $G_p^q$ and $Z_jG_p^q$ also works, though now the latter term is achieved by removing $j$ from the MS interaction.
 
 Compared to the technique from Ref.~\cite{casanova2012quantum}, our circuits once again cut the number of MS gates by half as for the regular single excitations. 

\subsection*{Simulation with Real-Valued Orbitals} \label{app:RealOrbitalSimulation}

For real orbitals, the one- and two-electron terms are real, thus simplifying the symmetries to $h_{pq} = h_{qp}$ and $h_{pqrs} = h_{qpsr} = h_{rspq} = h_{srqp}$. This changes the electronic Hamiltonian to
\begin{equation}
    H_{\mathrm{el.}} = \frac{1}{2}\sum_{pq} h_{pq} \tilde G_{p}^{q} + \frac{1}{4}\sum_{pqrs}h_{pqrs}\tilde G_{pq}^{rs},
\end{equation}
thus removing all the antisymmetric terms from equation~\ref{eq:AntisymmetrizedHamiltonian}.
At the same time, real orbitals introduce four additional permutation symmetries to the two-electron integrals, namely $h_{pqrs} = h_{rqps} = h_{spqr} = h_{psrq} = h_{qrsp}$ \cite{szabo_modern_1996}. This allows as to further simplify the Hamiltonian to
\begin{equation}
    H_{\mathrm{el.}} = \frac{1}{2}\sum_{pq}h_{pq} \tilde G_{p}^{q} + \frac{1}{8} \sum_{pqrs} h_{pqrs}\left(\tilde G_{pq}^{rs} + \tilde G_{ps}^{rq}\right).
\end{equation}
A derivation is provided in Supplementary Note \hyperref[app:RealOrbitals]{6}.
The term $\tilde G_{pq}^{rs} + \tilde G_{ps}^{rq}$ boils down to 4 Pauli strings instead of 8, which we can use to simplify the circuit structure. We use the antisymmetrized version $G_{pq}^{rs} + G_{ps}^{rq}$ to derive the corresponding circuit. From equation~\ref{eq:DoubleExcitationGeneratorPauli}, we conclude that
\begin{align}
    &G_{pq}^{rs} + G_{ps}^{rq} \to \frac{1}{4} \mathcal{Z}_{pq}^{rs} \\
    &\times (X_pX_qX_rY_s - X_pY_qX_rX_s + Y_pX_qY_rY_s - Y_pY_qY_rX_s). \nonumber
\end{align}
We can implement this term using the circuit from Fig.~\ref{fig:DoubleExcitation} by removing the $R_z$ gates on qubits $q_p$ and $q_r$ and adjusting the angles of the $R_z$ gates on qubits $q_q$ and $q_s$. Despite the reduction in Pauli strings, it still requires 4 MS gates. Hence, the four additional symmetries do not benefit the runtime of our quantum simulation scheme. 
However, assuming that $h_{pqrs}\neq h_{prsq}\neq 0$, the full 8 strings will be restored. 
Nonetheless, the use of real orbitals halves the number of terms in the Hamiltonian and followingly the depth of the Trotter step circuit. On a side note, linear combinations of the type $G_{pq}^{rs} \pm G_{ps}^{rq}$, also referred to as coupled exchange operators, have recently proven to be useful in variations of UCCSD theory \cite{ramoa2024reducing}.

\subsection*{The Experimental MS Gate} \label{subsec:MSGateEXP}

In the following, we focus on the \textbf{XX} gate with $\phi=0$ to simplify the notation. All considerations hold equivalently for the \textbf{YY} gate. The unitary operator corresponding to the experimental MS gate is obtained exactly by solving the time evolution operator via second-order Magnus expansion \cite{zhu2006arbitrary, Lotshaw2023Modeling}, and reads
\begin{align}
    \label{eq:MSGateExperiment}
    U_\text{MS}(t) &= \exp\left(-i\sum_{i<j}^N \chi^{(i,j)}(t) X_i X_j \right) \\
    &\phantom{=~} \times \exp\left(\sum_i^N \sum_{p}^N \left[\alpha_{p}^{(i)}(t) a_p^\dagger - \textrm{H.c.}\right] X_i  \right). \nonumber
\end{align}
The first part implements the spin-spin interaction between the ions electronic states/logical qubit states, whereas the second part captures the spin-displacement coupling between the electronic states and vibrational mode coherent states. Note that for $\chi^{(i,j)}=\theta/2$, the first term corresponds to the ideal MS gate assumed in equation~\ref{eq:MSGate}.  
The geometric phases defining the Ising evolution are given by
\begin{align} 
    \label{eq:geom_phases}
    \chi^{(i,j)}(t) &= \sum_p \eta_{p}^{(i)} \eta_{p}^{(j)} \int_0^t d\tau_2 \int_0^{\tau_2} d\tau_1 g^{(i)}(\tau_1) g^{(j)}(\tau_2) \nonumber \\ &\hphantom{~} \times \sin(\omega_p(\tau_2-\tau_1)), 
\end{align}
and the phase space displacements of the vibrational modes are
\begin{align}
    \alpha_{p}^{(i)}(t) = \int_0^t d\tau g^{(i)}(\tau) e^{i\omega_p \tau}. \label{eq:displacements}
\end{align}
By tracing out the phonon modes, one can quantify the decoherence from residual ion-vibration entanglement \cite{Lotshaw2023Modeling} and obtain the quantum channel acting solely on the qubit states.  

\subsection*{Pulse Modulation and Gate Stabilization} \label{subsec:PulseModulation}

In the following, we briefly recap the power-optimal stabilized entangling gate scheme from Ref.~\cite{blumel2021power}. A more detailed elaboration is provided in Supplementary Note~\hyperref[app:LinearQuadraticConstraints]{7}. The pulse functions (which entail both the amplitude and detuning) are expanded in a Fourier-sine basis. The pulse acting on the $i^{th}$ ion is defined as
\begin{equation}
    g^{(i)}(t)=\sum_{n=1}^{N_A} A_{n}^{(i)} \sin\left(2\pi n \frac{t}{T}\right)
    \label{eq:FourierPulse}
\end{equation}
where $T$ is the gate time, $A_{n}^{(i)}$ the $n^{th}$ Fourier coefficient and $N_A$ the total amount of Fourier coefficients. 
By construction, the pulse vanishes at $t=0$ and $t=T$. 
To ensure experimental viability of the pulses, the highest Fourier frequency is restricted to $\omega_{N_A}=2\pi\times N_A/T\leq2\pi\times\SI{5}{\mega\hertz}$, which is well within the capabilities of acusto-optical or electro-optical modulators typically used to modulate the laser beams.
To avoid residual ion-vibration entanglement, the pulses must satisfy the property $\alpha_{p}^{(i)}(T)=0$, which translates into $N$ linear constraints of the type $\sum_{n=1}^{N_A} M_{pn} A_{n}^{(i)} = 0$
with $M_{pn}=\int_0^t d\tau \sin{(2\pi n\tau/T)}e^{i\omega_p \tau}$.
To additionally stabilize this condition against vibrational mode fluctuations $\Delta \omega_p$ up to order $S$, the conditions $\partial^s/\partial \omega_p^s~\alpha_{p}^{(i)}(T)=0$ must be satisfied, which translates into $N\times S$ additional linear constraints $\sum_{n=1}^{N_A} M^{(s)}_{pn} A_{n}^{(i)} = 0$. 
By computing an orthonormal basis of the nullspace of $M$, one restricts the pulse-space basis to only such solutions which ensure vanishing residual ion-vibration entanglement, and reduces the dimension of our optimization problem. 
This then conveniently removes the second term of equation~\ref{eq:MSGateExperiment}. In our simulations, we use $S=10$ as we find that the dimension of the nullspace does not decrease beyond that.

Next, we require that all pair-wise entanglement degrees satisfy $\chi^{(i,j)}(T)=\pi/4$ for qubits $i$ and $j$ participating in the maximally entangling MS interaction. For a global MS interaction, this translates into $N\times (N-1)/2$ quadratic constraints of the type $\sum_{n,m=1}^{N_A} A_{n}^{(i)} S_{nm} A_{m}^{(j)} = \pi/4$ (cf.~Supplementary Note~\hyperref[app:LinearQuadraticConstraints]{7}). To make the entanglement degree robust against motional frequency fluctuations, one can again enforce vanishing derivatives up to some order $K$, which then yields $K\times N \times (N-1)/2$ additional quadratic constraints of the type $\sum_{n,m=1}^{N_A} A_{n}^{(i)} S^{(k)}_{nm} A_{j}^{(m)} = 0$. According to Ref.~\cite{blumel2021power}, the average laser power grows exponentially in $K$, which is why we choose the lowest stabilization order $K=1$ in our simulations.

The objective of finding power-optimal stabilized gates then boils down to minimizing the average laser power per ion $\gamma^2 \coloneqq 1/N \sum_i \sum_n (A_{n}^{(i)})^2$ under the quadratic constraints, which belongs to a class of optimization problems referred to as quadratically constrained quadratic program (QCQP) \cite{boyd2004convex}. 
For experimental application, optimizing the maximum instead of the average power would be more relevent. However, \cite{grzesiak2020efficient} show that both objectives lead to similar pulse shapes, with optimization on average power being more numerically stable.
For two-qubit MS gates, the globally optimal solution can be efficiently constructed \cite{grzesiak2020efficient, blumel2021power}. 
While the procedure does not generalize to global MS gates, it serves as a valuable initialization strategy in our simulations.

\clearpage

\section*{Acknowledgments}
This project was funded by the DLR Quantum Computing Initiative and the Federal Ministry for Economic Affairs and Climate Action; \url{https://qci.dlr.de/quanticom} (T.N.K, E.S, G.B.). 
We thank Max Haas for helpful comments on the manuscript. All quantum circuits were drawn using the \LaTeX-package \texttt{Quantikz} \cite{kay2018tutorial}. 

\section*{Data Availability}
The datasets generated and/or analysed during the current study are available in the Zenodo repository, \url{https://doi.org/10.5281/zenodo.18789051} \cite{Kaldenbach2025ImprovedStrategiesData}.

\section*{Code Availability}
The underlying code for this study is available in GitHub and can be accessed via this link \url{https://github.com/dlr-wf/MS-gate-noise-model}.

\section*{Competing interests}
A patent application filed by the German Aerospace Center (Deutsches Zentrum für Luft- und Raumfahrt e.V., DLR), currently pending with the German Patent and Trade Mark Office (Deutsches Patent- und Markenamt, DPMA), covers aspects of this work. It specifically includes the presented quantum circuits for symmetrized and antisymmetrized single-, controlled single\mbox{-,} and double-excitation operators and their applicability to UCCSD calculations and Hamiltonian simulation. The listed inventors are Thierry N. Kaldenbach, Gabriel Breuil, and Eric Breitbarth. The application number is DE 10 2025 133 101.4, with the German title "Quantenschaltkreis". The authors declare no other financial or non-financial competing interests.

\section*{Author Contributions}
T.N.K. developed the theoretical formalism, the simulation code, and performed the noisy simulations. T.N.K. and N.S. conceived the noise model. E.S. enabled the HPC integration of the simulation code. G.B. worked out the practical use cases in quantum chemistry and material science. T.N.K. lead the effort in writing the manuscript.

\printbibliography
\end{multicols*}

\clearpage

\supplementsetup
\section{Derivation of the Pauli Generator} \label{app:Generator}

In this Supplementary, we derive the expression for the Pauli generator $\mathcal{P}^{(j)}=\textbf{XX} Z_j\textbf{XX}^\dagger$, visualized in terms of the quantum circuit model. For an alternative derivation using trigonometric identities for the collective spin operator $S_x$, refer to Ref.~\cite{Müller2011Simulating}. 
First, we use that all the pairwise interactions $X_i X_k$ with $i, k \neq j$ do not affect $\mathcal{P}^{(j)}$:
\begin{equation}
    \mathcal{P}^{(j)}=\textbf{XX} Z_j \textbf{XX}^\dagger = \exp\Bigl(-i\frac{\pi}{4}\sum_{\substack{i \neq j}}X_iX_j\Bigr) Z_j \exp\Bigl(i\frac{\pi}{4}\sum_{\substack{i \neq j}}X_iX_j\Bigr).
\end{equation}
Next, we decompose the effectively remaining part of the \textbf{XX} gate into a sequence of $R_{xx}$ rotations
\begin{equation}
    \exp\Bigl(-i\frac{\pi}{4}\sum_{\substack{i \neq j}}X_iX_j\Bigr) = \prod_{i\neq j} \exp\Bigl(-i\frac{\pi}{4}X_iX_j\Bigr),
\end{equation}
where the circuit decomposition of the Clifford operation $\exp(-i\pi/4 X_iX_j)$ is given by:
\begin{equation}
    \vcenter{\hbox{\includegraphics{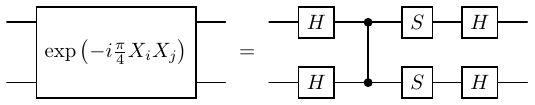}}}.
\end{equation}
Note that since $CZ$ and $S$ commute, we can capture all remaining $X_iX_j$ interactions with the following circuit:
\begin{equation}
   \vcenter{\hbox{\includegraphics{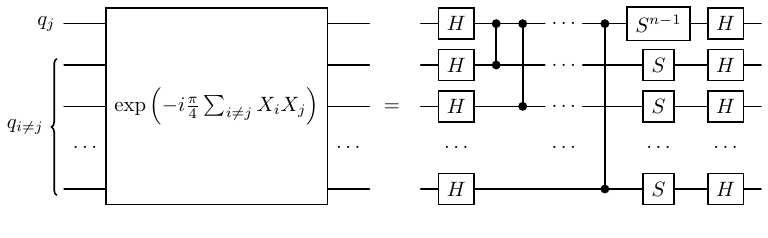}}}.
    \label{eq:MS_XX_IJ}
\end{equation}
We obtain $\mathcal{P}^{(j)}$ by propagating $Z_j$ through the circuit from equation~\ref{eq:MS_XX_IJ} from left to right (by the means of Heisenberg evolution).
\begin{equation}
    \mathcal{P}^{(j)} = \mathcal{X} \otimes (HS^{n-1}X S^{\dagger, n-1} H)_j ,
\end{equation}
where $\mathcal{X} = \bigotimes_{i\neq j}X_i$. Finally, by using that $S^2=Z$ and $S^3=S^\dagger$, as well as $SXS^\dagger = Y$, $S^\dagger X S = -Y$ and $ZXZ=-X$, and $H Y H = -Y$, we have
\begin{equation}
    \mathcal{P}^{(j)} = \mathcal{X} \otimes
    \begin{cases}
        \hphantom{-}Z_j, &\text{for } n=4k+1, k\in\mathbb{N}_0,\\
        -Y_j , &\text{for } n=4k+2, k\in\mathbb{N}_0,\\
        -Z_j, &\text{for } n=4k+3, k\in\mathbb{N}_0,\\
        \hphantom{-}Y_j , &\text{for } n=4k+4, k\in\mathbb{N}_0.\\
    \end{cases}
    \label{eq:GeneratorCases}
\end{equation}
Note that the above expression simplifies to $\mathcal{P}^{(j)}=Z_j$ for $n=1$, making it consistent with the case of a non-entangling single-qubit rotation. Equation~\ref{eq:Generator} in the main text is simply a short notation for equation~\ref{eq:GeneratorCases}.

\section{One- and Two-Electron Integrals in the Electronic Hamiltonian} \label{app:ElectronicHamiltonian}

The electronic structure Hamiltonian in second quantization reads 
\begin{equation}
    H_{\mathrm{el.}}=\sum_{pq}h_{pq}a_p^{\dagger}a_{q} + \frac{1}{2} \sum_{pqrs}h_{pqrs}a_{p}^{\dagger}a_{q}^{\dagger}a_{r}a_{s},
\end{equation}
where $h_{pq}$ and $h_{pqrs}$ are the one-electron and two-electron integrals, respectively. They can be efficiently classically computed for many different choices of basis functions $\phi_p(\mathbf{r})$ representing the orbitals. 
The one-electron integral $h_{pq}$ reads
\begin{equation}
    h_{pq}=\int d\mathbf{r}~\phi^{\ast}_{p}(\mathbf{r})\left(-\frac{1}{2}\nabla^2_{\mathbf{r}} -\sum_{I}\frac{Z_I}{||\mathbf{R}_I-\mathbf{r}||} + \frac{1}{2} \sum_{I}\sum_{J\neq I}\frac{Z_IZ_J}{||\mathbf{R}_I-\mathbf{R}_J||}\right)\phi_{q}(\mathbf{r}),
\end{equation}
where $Z_I$ is the atomic number of the $I$-th nucleus and $\bm{\mathrm{R}}_I$ its position. The one-electron integrals capture the electronic kinetic energy and the Coulomb interaction between electrons and nuclei, as well as between nuclei. Here, we neglect the kinetic energy of the nuclei  within the Born-Oppenheimer approximation.
The two-electron integral $h_{pqrs}$ is expressed as
\begin{equation}
 h_{pqrs}=\int d\mathbf{r}_1 d\mathbf{r}_2~ \frac{\phi^{\ast}_{p}(\mathbf{r}_1)\phi^{\ast}_{q}(\mathbf{r}_2)\phi_{r}(\mathbf{r}_1)\phi_{s}(\mathbf{r}_2)}{||\mathbf{r}_1-\mathbf{r}_2||},
\end{equation}
and corresponds to the electron-electron Coulomb repulsion.

\section{The Complex Electronic Hamiltonian} \label{app:ComplexOrbitals}

We simplify the quadratic fermionic terms by exploiting the permutational symmetry $h_{pq} = h^*_{qp}$:

\begin{align}
    \sum_{pq} h_{pq} a^\dagger_p a_q &= \frac{1}{2} \sum_{pq} \left[h_{pq} a^\dagger_p a_q + h_{qp} a^\dagger_q a_p\right] \nonumber \\
    &= \frac{1}{2} \sum_{pq} \left[h_{pq} a^\dagger_p a_q + h^*_{pq} a^\dagger_q a_p\right] \nonumber\\
    &= \frac{1}{2} \sum_{pq} \left[\Re(h_{pq}) \left(a^\dagger_p a_q + a^\dagger_q a_p\right) + i\Im(h_{pq}) \left(a^\dagger_p a_q - a^\dagger_q a_p\right)\right] \nonumber\\
    &= \frac{1}{2} \sum_{pq} \left[\Re(h_{pq}) \tilde G_p^q + \Im(h_{pq}) G_p^q\right].
\end{align}
Using that $h^*_{pp} = h_{pp}$, we rewrite the result as follows:
\begin{equation}
    \sum_{pq} h_{pq} a^\dagger_p a_q = \frac{1}{2} \sum_p h_{pp} \tilde G_p^p + \sum_{p<q} \left[\Re(h_{pq}) \tilde G_p^q + \Im(h_{pq}) G_p^q\right]
\end{equation}
Next, we simplify the quartic fermionic terms by exploiting the following permutational symmetries for complex orbitals:

\begin{equation}
    h_{pqrs} = h_{qpsr} = h^*_{rspq} = h^*_{srqp}.
\end{equation}

\begin{align}
    \frac{1}{2} \sum_{pqrs}h_{pqrs}a_{p}^{\dagger}a_{q}^{\dagger}a_{r}a_{s} &= \frac{1}{8} \sum_{pqrs} \left(
    h_{pqrs}a_{p}^{\dagger}a_{q}^{\dagger}a_{r}a_{s} +
    h_{qpsr}a_{q}^{\dagger}a_{p}^{\dagger}a_{s}a_{r} + 
    h_{rspq}a_{r}^{\dagger}a_{s}^{\dagger}a_{p}a_{q} + 
    h_{srqp}a_{s}^{\dagger}a_{r}^{\dagger}a_{q}a_{p}
    \right) \nonumber\\
    &= \frac{1}{8} \sum_{pqrs} \left(
    h_{pqrs}a_{p}^{\dagger}a_{q}^{\dagger}a_{r}a_{s} +
    h_{pqrs}a_{q}^{\dagger}a_{p}^{\dagger}a_{s}a_{r} + 
    h^*_{pqrs}a_{r}^{\dagger}a_{s}^{\dagger}a_{p}a_{q} + 
    h^*_{pqrs}a_{s}^{\dagger}a_{r}^{\dagger}a_{q}a_{p}
    \right) \nonumber\\
    &= \frac{1}{4} \sum_{pqrs} \left(
    h_{pqrs}a_{p}^{\dagger}a_{q}^{\dagger}a_{r}a_{s} +
    h^*_{pqrs}a_{r}^{\dagger}a_{s}^{\dagger}a_{p}a_{q} 
    \right) \nonumber\\
    &= \frac{1}{4} \sum_{pqrs} \left[
    \Re(h_{pqrs})\left(a_{p}^{\dagger}a_{q}^{\dagger}a_{r}a_{s} +a_{r}^{\dagger}a_{s}^{\dagger}a_{p}a_{q} \right)
    + i \Im(h_{pqrs})\left(a_{p}^{\dagger}a_{q}^{\dagger}a_{r}a_{s} - a_{r}^{\dagger}a_{s}^{\dagger}a_{p}a_{q} \right)
    \right] \nonumber\\
    &= \frac{1}{4} \sum_{pqrs} \left[\Re(h_{pqrs}) \tilde G_{pq}^{rs} + \Im(h_{pqrs}) G_{pq}^{rs}\right].
\end{align}

\section{Decomposition with Two Global MS Gates} \label{app:CNOT}

In this Appendix, we want to shortly address how the global MS gate can be paired with other 2-local Clifford entangling gates, such as the two-qubit MS gate $R_{XX}(\pi/2)$, the controlled phase gate $CZ$, or the controlled NOT gate $CX$, to reduce the number of global MS gates to 2 regardless of the excitation order. One could obviously argue that any MS gate may be decomposed as a sequence of two-qubits gates, but that is not the point here. Instead, we seek a decomposition where the MS gates account for the non-locality of the parity strings $\mathcal{Z}$ in the JW mapping, while all other entanglement on the orbitals subject to the excitation shall be captured by 2-local entangling gates. 

Let us once again consider the $XYYY$ strings. We can decompose this string as the product $XYYY = - XYXX\cdot XXYX \cdot XXXY$. At the same time, we already know how to implement this product, namely
\begin{align}
    &\textbf{XX} (Z_q Z_r Z_s) \textbf{XX}^\dagger = \textbf{XX} Z_q \textbf{XX}^\dagger \textbf{XX} Z_r \textbf{XX}^\dagger \textbf{XX} Z_s \textbf{XX}^\dagger \nonumber\\
    &= (-1)^m \mathcal{X}_{pq}^{rs} X_pY_qX_rX_s \cdot X_pX_qY_rX_s \cdot X_pX_qX_rY_s \nonumber\\
    &= (-1)^m \mathcal{X}_{pq}^{rs}(-X_pY_qY_rY_s).
    \label{eq:XYYYStringDecomposition}
\end{align}
As a consequence, we can also implement the $XYYY$ strings with \textbf{XX} gates by using 3-local rotations $\exp(-i\theta/2 ZZZ)$. These can be trivially decomposed into $R_z$ and CNOT gates. Finally, we can capture all $XYYY$ strings of the double excitation using
\begin{align}
    &\textbf{XX} (-Z_q Z_r Z_s - Z_p Z_r Z_s + Z_p Z_q Z_s + Z_p Z_q Z_r)\textbf{XX}^\dagger \nonumber\\
    &= (-1)^m \mathcal{X}_{pq}^{rs}(X_pY_qY_rY_s + Y_pX_qY_rY_s - Y_pY_qX_rY_s - Y_pY_qY_rX_s).  \label{eq:DoubleExcitationGeneratorDecompositionYYYXCX} 
\end{align}
Combining this with equation~\ref{eq:DoubleExcitationGeneratorDecompositionXXXYEven}, we can implement the double excitation from equation~\ref{eq:DoubleExcitationGeneratorPauli} with only 2 MS gates at the expense of introducing four non-local Pauli rotations. These rotations are however local to the qubits $p,q,r,s$ directly affected by the excitation. The scheme can be readily generalized to arbitrary excitations. 

\section{Equivalence of Symmetrized- \\and Antisymmetrized Excitations} \label{app:Equivalence}

In this Appendix, we derive the local equivalence of symmetrized- and antisymmetrized excitation operators in fermionic space. For that purpose, we first consider the following unitary:
\begin{equation}
    U=\exp(-i\theta n_j) \overset{n_j^2=n_j}{=} 1 + n_j \sum_{k=1}^\infty \frac{(-i\theta)^k}{k!} = 1 + (\exp(-i\theta) - 1) n_j. 
\end{equation}
Next, we conjugate the fermionic creation operator with this unitary.
\begin{align}
    \exp(-i\theta n_j) a_j^{\dagger} \exp(i\theta n_j) &= \exp(-i\theta n_j) a_j^{\dagger} \left[1 + (\exp(i\theta) - 1) n_j\right] \nonumber\\
    &= \exp(-i\theta n_j) \Bigl[a_j^\dagger + (\exp(i\theta) - 1) \underbrace{a_j^\dagger a_j^\dagger a_j}_{=0}\Bigr] \nonumber\\
    &= \left[1 + (\exp(-i\theta) - 1) n_j\right] a_j^\dagger \nonumber\\
    &= a_j^\dagger + (\exp(-i\theta) - 1) a_j^\dagger a_j a_j^\dagger \nonumber\\
    &= a_j^\dagger + (\exp(-i\theta) - 1) a_j^\dagger \nonumber\\
    &= \exp(-i\theta) a_j^\dagger. 
\end{align}
In the same manner (or simply through Hermitian conjugation), for the fermionic annihilation operator we obtain
\begin{equation}
     \exp(-i\theta n_j) a_j \exp(i\theta n_j) = \exp(i\theta) a_j. 
\end{equation}
Note that the induced phases differ by a sign. We exploit this property at $\theta = \pi/2$ to multiply the creation operators by $+i$ and the annihilation operators by $-i$:
\begin{equation}
    \exp\left(-i\frac{\pi}{2} n_j\right) a_j^{(\dagger)} \exp\left(i\frac{\pi}{2} n_j\right) = \varpm{} i a_j^{(\dagger)}.
\end{equation}
We now use this property to derive the equivalence between single excitations and quadratic Hamiltonian terms.
\begin{align}
    \exp\left(-i\frac{\pi}{2}n_p\right) G_p^q \exp\left(i\frac{\pi}{2}n_p\right) &= i \exp\left(-i\frac{\pi}{2}n_p\right) \left(a^\dagger_p a_q - a^\dagger_q a_p \right) \exp\left(i\frac{\pi}{2}n_p\right) \nonumber\\
    &= i \left(-i a^\dagger_q a_p - i a^\dagger_p a_q\right) \nonumber\\
    &= \left(a^\dagger_q a_p + a^\dagger_p a_q\right) \nonumber\\
    &= \tilde G_p^q.
\end{align}
If we do the conjugation on the fermionic mode $q$ instead of $p$, we obtain
\begin{equation}
    \exp\left(-i\frac{\pi}{2}n_q\right) G_p^q \exp\left(i\frac{\pi}{2}n_q\right) = - \tilde G_p^q.
\end{equation}
Finally, we have
\begin{equation}
    \exp\left(-i\frac{\pi}{2}n_j\right) G_p^q \exp\left(i\frac{\pi}{2}n_j\right) = 
    \begin{cases}
        +\tilde G_p^q   & \text{if } j=p, \\
        -\tilde G_p^q   & \text{if } j=q, \\
        +       G_p^q   & \text{else}.
    \end{cases}
\end{equation}
For the double excitations and quartic Hamiltonian terms, we analogously find
\begin{equation}
    \exp\left(-i\frac{\pi}{2}n_j\right) G_{pq}^{rs} \exp\left(i\frac{\pi}{2}n_j\right) = 
    \begin{cases}
        +\tilde G_{pq}^{rs}   & \text{if } j=p \text{ or } j=q, \\
        -\tilde G_{pq}^{rs}   & \text{if } j=r \text{ or } j=s, \\
        +       G_{pq}^{rs}   & \text{else}.
    \end{cases}
\end{equation}

\section{The Real Electronic Hamiltonian} \label{app:RealOrbitals}

Using that all integrals are real for a real basis, we simplify the quadratic terms to 
\begin{align}
    \sum_{pq} h_{pq} a^\dagger_p a_q &= \frac{1}{2} \sum_{pq} h_{pq}\tilde G_p^q  \nonumber\\ 
    &= \frac{1}{2} \sum_p h_{pp} \tilde G_p^p + \sum_{p<q} h_{pq} \tilde G_p^q,
\end{align}
and the quartic terms to
\begin{align}
    \frac{1}{2} \sum_{pqrs}h_{pqrs}a_{p}^{\dagger}a_{q}^{\dagger}a_{r}a_{s} = \frac{1}{4} \sum_{pqrs} h_{pqrs} \tilde G_{pq}^{rs}.
\end{align}
Employing all the symmetries
\begin{equation}
    h_{pqrs} = h_{qpsr} = h_{rspq} = h_{srqp} = h_{rqps} = h_{spqr} = h_{psrq} = h_{qrsp},
\end{equation}
gives rise to
\begin{align}
    \frac{1}{4} \sum_{pqrs} h_{pqrs} \tilde G_{pq}^{rs} &=\frac{1}{32} \sum_{pqrs} \left[
    h_{pqrs} \tilde G_{pq}^{rs} +
    h_{qpsr} \tilde G_{qp}^{sr} +
    h_{rspq} \tilde G_{rs}^{pq} +
    h_{srqp} \tilde G_{sr}^{pq} +
    h_{rqps} \tilde G_{rq}^{ps} + 
    h_{spqr} \tilde G_{sp}^{qr} + 
    h_{psrq} \tilde G_{ps}^{rq} + 
    h_{qrsp} \tilde G_{qr}^{sp} \right] \nonumber \\
    &= \frac{1}{32} \sum_{pqrs} h_{pqrs} \left[
    \tilde G_{pq}^{rs} + 
    \tilde G_{qp}^{sr} +
    \tilde G_{rs}^{pq} +
    \tilde G_{sr}^{qp} +
    \tilde G_{rq}^{ps} + 
    \tilde G_{sp}^{qr} + 
    \tilde G_{ps}^{rq} + 
    \tilde G_{qr}^{sp} \right] \nonumber\\
    &= \frac{1}{8} \sum_{pqrs} h_{pqrs} \left[\tilde G_{pq}^{rs} + \tilde G_{ps}^{rq} \right].
\end{align}

\section{Pulse Modulation as a QCQP} \label{app:LinearQuadraticConstraints}

This appendix is based on the power-optimal pulse stabilization method provided in Ref.~\cite{blumel2021power}. We summarize the approach, and supplement it with some analytical solutions.

We assume we have $P$ vibrational modes and we set the number of Fourier coefficients $N_A > P$ in order to have non-trivial solutions for the linear and quadratic constraints. In addition, we assume that all laser pulses have negative parity, which ensures that the pulse functions vanish at $t=0$ and $t=T$. 

\paragraph{Linear constraints:} We are first concerned with shaping the pulses such that the residual entanglement between electronic and vibrational states vanishes at the end of the gate, i.e.~$\alpha_{p}^{(i)}(T)=0$. To ensure that this property holds in presence of motional mode frequency fluctuations $\Delta \omega_p$, we additionally enforce frequency stabilizations up to some order $S$, which in total we can write as $\partial^s/\partial\omega_p^s~\alpha_{p}^{(i)}(T)=0$ for all $0 \leq s\leq S$.

More explicitly, these properties can be accomplished by considering the following $P(S+1)$ linear constraints:
\begin{align}
    \frac{\partial^s\alpha_{p}^{(i)}(T)}{\partial\omega_p^s}=\frac{\partial^s}{\partial\omega_p^s}\sum_{n=1}^{N_A} A_{n}^{(i)} \int_0^T dt\sin\left(2\pi n \frac{t}{T}\right)\sin\left[\omega_p \left(\frac{T}{2}-t\right)\right]=0.
    \label{eq:LinearConstraint}
\end{align}
We note that the constraints are effectively the same for the pulses of all ions, and therefore omit the ion index $i$ for now. It is convenient to first carry out the integral
\begin{align}
    \int_0^T dt\sin\left(2n\pi \frac{t}{T}\right)\sin\left[\omega_p \left(\frac{T}{2}-t\right)\right]= \frac{T}{\pi}\sin\left(\omega_p\frac{ T}{2}\right)\frac{n}{n^2-n_p^2},
\end{align}
where $n_p \coloneqq \omega_p T/(2\pi)$ is the number of oscillations of the vibrational mode $p$ during the gate. It is then clear that the stabilization of order $s=0$ results in the $P$ linear equations
\begin{align}
    \sum_{n=1}^{N_A} A_n \frac{n}{n^2-n_p^2} = 0.
\end{align}
For the higher order stabilizations ($s>0$), one needs to carry out the partial derivatives w.r.t.~$\omega_p$. It is then straightforward to show that the following additional conditions must be satisfied:
\begin{align}
    \sum_{n=1}^{N_A} A_n \frac{n}{(n^2-n_p^2)^{s+1}} = 0
\end{align}
Last, it is convenient to rewrite these constraints as a matrix-vector product
\begin{align}
    M\Vec{A}=0,
\end{align}
where $M$ is a $P(S+1)\times N_A$ matrix, and $\Vec{A}$ is the amplitude vector gathering the Fourier coefficients of the pulse function of \eqref{eq:FourierPulse}. A basis of the stabilized pulse-space is then obtained by finding the nullspace of $M$, which can be equivalently formulated as finding the eigenvectors $\vec A^{(\alpha)}$ of $\Gamma = M^T M$ with eigenvalues $0$. Any stabilized pulse can then be expanded in terms of these eigenvectors as
\begin{align}
    \vec A = \sum_{\alpha=1}^{N_0} \Lambda_\alpha \vec A^{(\alpha)},
\end{align}
where $N_0 < N_A$ is the dimension of the nullspace.

\paragraph{Quadratic constraints:} To achieve the $N$-qubit MS gate with $\theta=\pi/2$, the geometric phases must satisfy the condition $\chi^{(i, j)}(T)=\pi/8$ on all ions participating in the gate. This is translated into $N(N-1)/2$ quadratic constraints as follows:
\begin{align}
    \chi^{(i,j)}(T) = \sum_{n=1}^{N_A} \sum_{m=1}^{N_A} A_{n}^{(i)} A_{m}^{(j)} \sum_{p=1}^P \eta_{p}^{(i)} \eta_{p}^{(j)}
    \int_0^Td\tau_2\int_0^{\tau_2}d\tau_1
    \sin\left(2\pi n\frac{\tau_2}{T}\right)
    \sin\left(2\pi m\frac{\tau_1}{T}\right)
    \sin{\left[\omega_p \left( \tau_2-\tau_1\right)\right]} \textcolor{brown}{\overset{!}{=}} \frac{\pi}{8}
\end{align}
The double integral can be exactly solved as
\begin{align}
    \int_0^Td\tau_2\int_0^{\tau_2}d\tau_1
    \sin\left(2\pi n\frac{\tau_2}{T}\right)
    \sin\left(2\pi m\frac{\tau_1}{T}\right)
    \sin{\left[\omega_p \left( \tau_2-\tau_1\right)\right]} = - \left(\frac{T}{2\pi}\right)^2
    \begin{cases}
        \frac{nm\sin(\omega_p T)}{(n^2-n_p^2)(m^2-n_p^2)} & \text{if } n \neq m, \\ 
        \frac{n^2\sin(\omega_p T)}{(n^2-n_p^2)^2} + \frac{1}{2} \frac{\omega_p T}{n^2-n_p^2} & \text{if } n=m.
    \end{cases}
\end{align}
In analogy to the matrix-vector formalism for the linear constraints, one can then write $\chi^{(i,j)}(T) = \sum_n\sum_m A_{n}^{(i)} A_{m}^{(j)} S^{(i,j)}_{nm}$, or in short 
\begin{align}
    \chi^{(i,j)}(T) = \vec A^{(i),T} S^{(i, j)} \vec A^{(j)}  \textcolor{brown}{\overset{!}{=}} \frac{\pi}{8}. 
\end{align}
In this form, it is straightforward to project the quadratic constraint onto the subspace which satisfies the linear constraints, which then yields
\begin{align}
    \chi^{(i,j)}(T) = \vec \Lambda^{(i),T} R^{(i, j)} \vec \Lambda^{(j)} = \frac{\pi}{8}, 
\end{align}
where $R$ is the symmetrized projection of $D$ onto the nullspace basis of $M$. To stabilize against motional frequency fluctuations up to order $K$, the additional constraints $\partial^k/\partial \omega_p^k~\chi^{(i,j)} = 0$ must be satisfied for all $0 < k \leq K$, which results in $P K N(N-1)/2$ additional quadratic constraints. The exact solutions for $\partial^k/\partial \omega_p^k~R^{(i,j)}$ are tedious, but can in principle be computed efficiently by rewriting the double integral solution in terms of partial fractions, thus avoiding numerical differentiation.

\paragraph{Quadratic Objective:} To find experimentally feasible solutions, the pulses are optimized w.r.t. to the average laser power throughout the gate. This results in a quadratic objective function
\begin{align}
    \gamma^2 = \frac{2}{T} \sum_i \int_0^T dt \left[g^{(i)}(\tau)\right]^2 = \sum_i \sum_{n=1}^{N_A} {A_n^{(i)}}^2 = \sum_i \sum_{\alpha=1}^{N_0} {\Lambda_\alpha^{(i)}}^2.
\end{align}
In summary, the linear constraints are directly encoded into the program, such that only a quadratic objective and quadratic constraints remain, which define the QCQP.

\section{Noise Model Characterization} \label{app:NoiseModel}

In this section, we provide additional details on the employed noise model. Based on the axial trap frequency $\omega_z = 2\pi \times \SI{0.5}{\mega \hertz}$ and radial trap frequency $\omega_x = 2\pi \times  \SI{3.33}{\mega\hertz}$, we start by computing the radial motional modes frequencies $\omega_p$ and normal mode transformation matrices $b_p^{(i)}$, which we need for both linear and quadratic constraints in the noise model. Following, e.g., Ref.~\cite{Wu2018Noise}, we compute the values summarized in Supplementary Table~\ref{tab:MotionalModes}.
\begin{table}[h!]
    \centering
    \begin{adjustbox}{width=\textwidth}
    \begin{tabular}{l|c|cccccccccccc}
                & $\omega_p/2\pi$ [\si{\mega\hertz}] 
                & $i=1$ & $i=2$ & $i=3$ & $i=4$ & $i=5$ & $i=6$ & $i=7$ & $i=8$ & $i=9$ & $i=10$ & $i=11$ & $i=12$\\ \hline
        $p=1$   & 1.9646    
                & 0.000 & -0.006 & 0.044 & -0.166 & 0.381 & -0.570 & 0.570 & -0.381 & 0.166 & -0.044 & 0.006 & -0.000 \\

        $p=2$   & 2.2048    
                & 0.002 & -0.027 & 0.144 & -0.381 & 0.518 & -0.256 & -0.256 & 0.518 & -0.381 & 0.144 & -0.027 & 0.002 \\

        $p=3$   & 2.4087    
                & -0.007 & 0.079 & -0.303 & 0.498 & -0.226 & -0.320 & 0.320 & 0.226 & -0.498 & 0.303 & -0.079 & 0.007 \\

        $p=4$   & 2.5846    
                & 0.021 & -0.176 & 0.453 & -0.344 & -0.245 & 0.291 & 0.291 & -0.245 & -0.344 & 0.453 & -0.176 & 0.021 \\

        $p=5$   & 2.7375    
                & 0.054 & -0.313 & 0.473 & 0.040 & -0.359 & -0.212 & 0.212 & 0.359 & -0.040 & -0.473 & 0.313 & -0.054 \\

        $p=6$   & 2.8709    
                & 0.116 & -0.445 & 0.279 & 0.346 & 0.004 & -0.300 & -0.300 & 0.004 & 0.346 & 0.279 & -0.445 & 0.116 \\

        $p=7$   & 2.9867    
                & -0.215 & 0.494 & 0.064 & -0.280 & -0.328 & -0.138 & 0.138 & 0.328 & 0.280 & -0.064 & -0.494 & 0.215 \\

        $p=8$   & 3.0866    
                & 0.343 & -0.385 & -0.334 & -0.083 & 0.160 & 0.300 & 0.300 & 0.160 & -0.083 & -0.334 & -0.385 & 0.343 \\

        $p=9$   & 3.1713    
                & 0.468 & -0.118 & -0.316 & -0.326 & -0.232 & -0.083 & 0.083 & 0.232 & 0.326 & 0.316 & 0.118 & -0.468 \\

        $p=10$  & 3.2411    
                & 0.534 & 0.195 & -0.020 & -0.163 & -0.252 & -0.294 & -0.294 & -0.252 & -0.163 & -0.020 & 0.195 & 0.534 \\

        $p=11$  & 3.2955    
                & -0.482 & -0.371 & -0.279 & -0.195 & -0.115 & -0.038 & 0.038 & 0.115 & 0.195 & 0.279 & 0.371 & 0.482 \\

        $p=12$  & 3.3333    
                & 0.289 & 0.289 & 0.289 & 0.289 & 0.289 & 0.289 & 0.289 & 0.289 & 0.289 & 0.289 & 0.289 & 0.289 \\       
    \end{tabular}
    \end{adjustbox}
    \caption{Radial normal mode frequencies $\omega_p$ and normal mode transformation matrix coefficients $b_p^{(i)}$.}
    \label{tab:MotionalModes}
\end{table}

For our simulations, we assume that the 12-qubit MS gate has a gate time of $\SI{300}{\micro\s}$, and we model each pulse by $N_A=1500$ Fourier components, such that the highest frequency is $\SI{5}{\mega\hertz}$.
Thus, the modulation speeds required for the resulting pulses are well within the bandwidths of acusto-optical or electro-optical modulators typically used to control the laser beams. Also, the smallest spectral gap is resolved by 10 Fourier components, which we observe to be sufficient for numerical stability here.
We stabilize the displacements against frequency fluctuations $\Delta \omega_p$ up to order $S=10$, since beyond that we do not observe reductions in the nullspace dimension. For the entanglement degrees, we first run an optimization without stabilization ($K=0$), which we then use as an initial guess for the optimization with first-order stabilization $K=1$. We find that this significantly speeds up finding the solution compared to greedily starting with $K=1$. The power-optimal pulse functions for each of the 12 ions are shown in Supplementary Fig.~\ref{fig:pulses}. 
\begin{figure}[t]
    \centering
    \includegraphics[width=\textwidth]{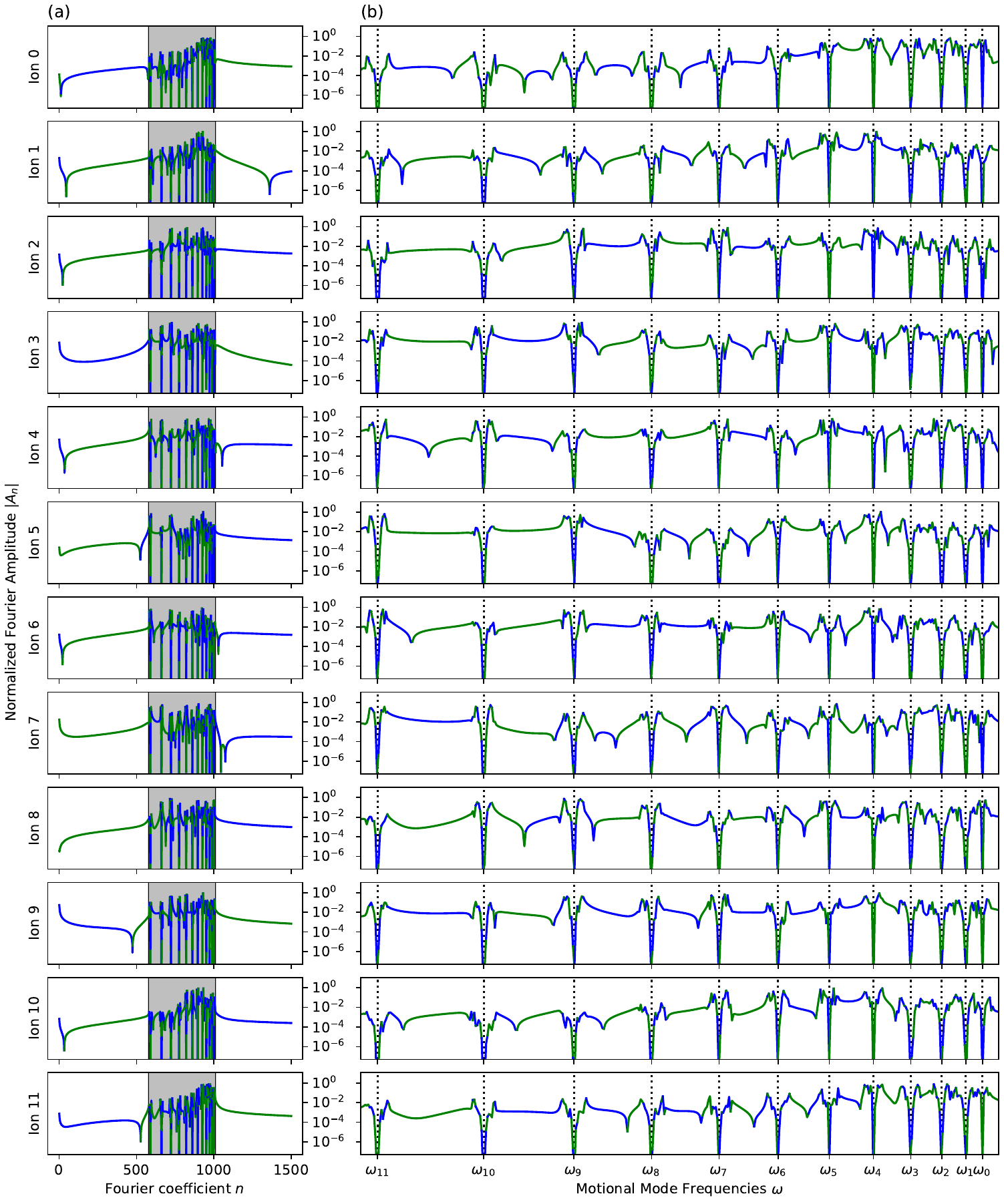}
    \caption{Power-optimal pulse functions $g^{(i)}$ for a global MS gate in the 12 qubit ion trap. (a) The normalized Fourier-sine coefficients $|A^{(i)}_n|$ in the full frequency range, (b) a close-up on the normal mode frequency range. The sign of th $A^{(i)}_N$ is displayed in color coding, with $A^{(i)}_N\gtrless0$ being indicated in blue/green. The pulses vanish around every normal mode frequency $\omega_p$, thus avoiding heating of the modes in the presence of fluctuations. We note that this is fundamentally different from the two-qubit MS gates in \cite{blumel2021power}, where optimal pulses target the resonances.}
    \label{fig:pulses}
\end{figure}

\begin{figure}
    \centering
    \includegraphics[width=0.8\textwidth]{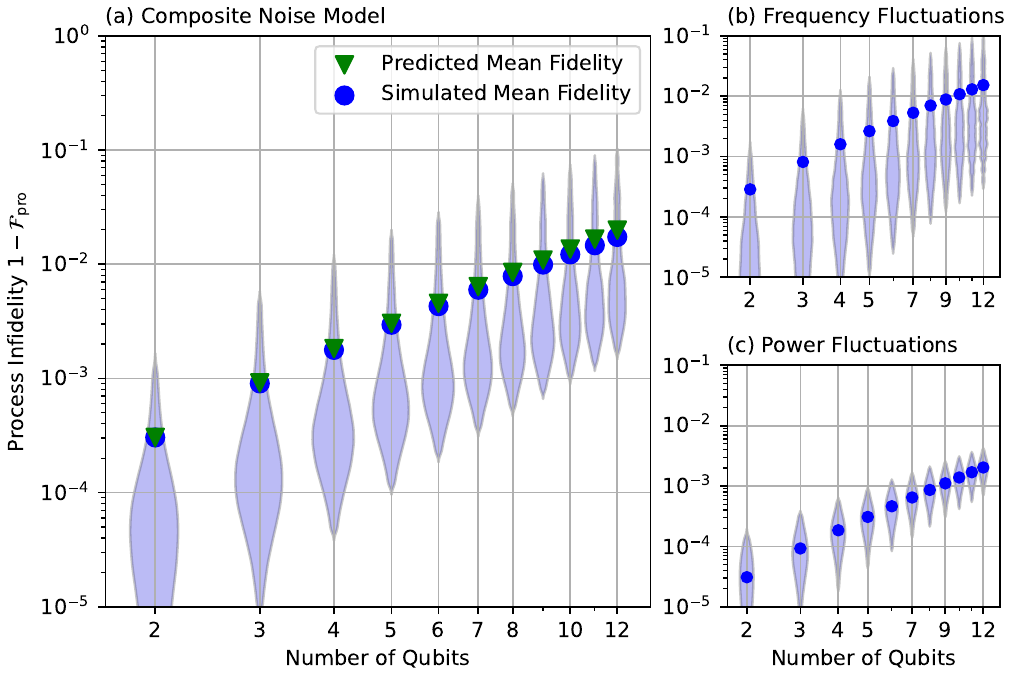}
    \caption{Benchmarking of the infidelity of a single MS gate. Note the double-logarithmic axes. The blue violin plots depict $3\sigma$ of the infidelity distribution over \num{1000000} unique realizations of frequency- and power fluctuations for each locality, while the blue circle marks the average infidelity, and the green triangle marks a simple infidelity prediction based on exponentiating the fidelity of a two-qubit gate with the number of pairwise two-qubit gates entailed in an $N$-qubit MS gate. (a) The composite noise model including both frequency and power fluctuations. The isolated impacts of frequency- and power fluctuations are shown in (b) and (c), respectively. We assume frequency fluctuations of \SI{1}{\kilo\hertz} and power fluctuations of \SI{0.5}{\percent}.}
    \label{fig:MSGATE_BENCHMARK}
\end{figure}
We now characterize the system by using these pulses to compute the gate infidelities. For the sake of simplicity, when targeting subsets of ions on the trap, we simply turn off the pulses on the respective ions. While this no longer presents power-optimal pulses (we come back to this later), and thus overestimates laser-power fluctuations, we consider it sufficient to assess the noise. Assuming frequency fluctuations of $\SI{1}{\kilo\hertz}$ \cite{blumel2021power} and power fluctuations of $\SI{0.5}{\%}$, we obtain the results shown in Supplementary Fig.~\ref{fig:MSGATE_BENCHMARK} (a). 
We find a two-qubit gate fidelity of $\mathcal{F}_2=\SI{99.97}{\%}$, and observe an exponential decay in fidelity when increasing the number of qubits, up to a fidelity of $\mathcal{F}_{12}=\SI{98.27}{\%}$. We find that the decay is well described by $\mathcal{F}_2^{N(N-1)/2}$, which is motivated by the fact that an $N$-qubit MS gate implicitly entails $N(N-1)/2$ two-qubit MS gates.
Supplementary Figs.~\ref{fig:MSGATE_BENCHMARK} (b) and (c) show the infidelity when we only consider trap frequency or power fluctuation, respectively. 
We find that fluctuations of the trap frequency is the main source of infidelity even after applying the frequency stabilization formalism from Ref.~\cite{blumel2021power}, which is why applying stabilization formalism is crucial for our work.

Finally, in Supplementary Fig.~\ref{fig:PULSETIME} we investigate how the MS gate time scales in our assumed setup. For that, we conduct two types of experiments. First, we consider a fixed gate time of $\SI{300}{\micro\s}$ regardless of the number of qubits, and compute the average laser power per ion (Supplementary Fig.~\ref{fig:PULSETIME} (a)).
We find that the average laser power per qubit decreases when we reduce the number of qubits in the MS gate. This suggests that smaller gates should be realized with faster pulses to adequately utilize the available laser power. 
Therefore, second, we consider a linear gate time scaling, where the two-qubit MS gate takes $\SI{50}{\micro\s}$, and the 12-qubit MS gate is $\SI{300}{\micro\s}$ as previously (Supplementary Fig.~\ref{fig:PULSETIME} (b)). 
Here, we find that the gate times are too short, leading to higher average powers per ion for smaller gates. 
Assessing a precise scaling behavior of optimal gate time is not possible within our limited data. 
However, it is clear that the gate time must be sub-linear, which counteracts the issue of linear locality of fermionic simulations within the JW mapping. 

\begin{figure}
    \centering
    \includegraphics[width=0.8\textwidth]{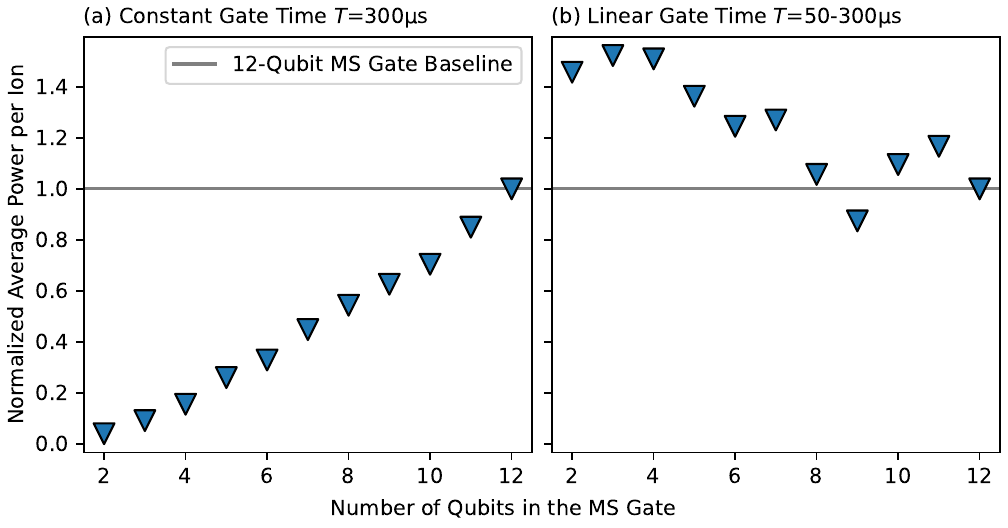}
    \caption{Average laser powers per ion in targeted MS gates normalized with respect to the 12-qubit MS gate. (a) The average power assuming a fixed gate time of $\SI{300}{\micro\s}$ regardless of the number of qubits. (b) The average power assuming a linear gate time ranging from $\SI{50}{\micro\s}$ for two-qubit MS gate takes to $\SI{300}{\micro\s}$ for the  12-qubit MS gate.}
    \label{fig:PULSETIME}
\end{figure}

\end{document}